
\newcount\mgnf\newcount\tipi\newcount\tipoformule\newcount\driver
\newcount\indice\newcount\tipoeq

\driver=1        
\mgnf=0          
\tipi=2          
\tipoformule=0   
\indice=1        
\tipoeq=1        
\ifnum\mgnf=0
   \magnification=\magstep0
   \voffset=0.truecm\hsize=14.5truecm\vsize=23.truecm
   \parindent=4.pt\baselineskip=0.45cm\fi
\ifnum\mgnf=1
   \magnification=\magstep1
\hsize=14.5truecm\vsize=23.truecm
   \baselineskip=0.5cm plus0.1pt minus0.1pt \parindent=0.9truecm
   \lineskip=0.5truecm\lineskiplimit=0.1pt      \parskip=0.1pt plus1pt\fi
\let\a=\alpha \let\b=\beta  \let\g=\gamma     \let\d=\delta  \let\e=\varepsilon
  \let\h=\eta   \let\th=\vartheta \let\k=\kappa   \let\l=\lambda
\let\m=\mu    \let\n=\nu    \let\x=\xi        \let\p=\pi      
\let\s=\sigma \let\t=\tau   \let\f=\varphi     
               \let\o=\omega

\let\G=\Gamma \let\D=\Delta     \let\L=\Lambda  
    \let\Si=\Sigma \let\F=\Phi

\global\newcount\numsec\global\newcount\numfor
\global\newcount\numapp\global\newcount\numcap
\global\newcount\numfig\global\newcount\numpag
\global\newcount\numnf

\def\SIA #1,#2,#3 {\senondefinito{#1#2}%
\expandafter\xdef\csname #1#2\endcsname{#3}\else
\write16{???? ma #1,#2 e' gia' stato definito !!!!} \fi}

\def \FU(#1)#2{\SIA fu,#1,#2 }

\ifnum\tipoeq=1
\def\etichetta(#1){(\veroparagrafo.\veraformula)%
\SIA e,#1,(\veroparagrafo.\veraformula) %
\global\advance\numfor by 1%
\write15{\string\FU (#1){\equ(#1)}}%
\write16{ EQ #1 ==> \equ(#1)  }}
\def\etichettaa(#1){(A\veraappendice.\veraformula)
 \SIA e,#1,(A\veraappendice.\veraformula)
 \global\advance\numfor by 1
 \write15{\string\FU (#1){\equ(#1)}}
 \write16{ EQ #1 ==> \equ(#1) }}
\def\getichetta(#1){Fig. \verafigura
 \SIA g,#1,{\verafigura}
 \global\advance\numfig by 1
 \write15{\string\FU (#1){\graf(#1)}}
 \write16{ Fig. #1 ==> \graf(#1) }}
\def\retichetta(#1){\numpag=\pgn\SIA r,#1,{\verapagina}
 \write15{\string\FU (#1){\rif(#1)}}
 \write16{\rif(#1) ha simbolo  #1  }}
\def\etichettan(#1){(n\verocapitolo.\veranformula)
 \SIA e,#1,(n\verocapitolo.\veranformula)
 \global\advance\numnf by 1
\write16{\equ(#1) <= #1  }}
\fi

\ifnum\tipoeq=0
\def\etichetta(#1){(\veraformula)
\SIA e,#1,(\veraformula)
\global\advance\numfor by 1
\write15{\string\FU (#1){\equ(#1)}}
\write16{ EQ #1 ==> \equ(#1)  }}
\def\etichettaa(#1){(A\veraformula)
 \SIA e,#1,(A\veraformula)
 \global\advance\numfor by 1
 \write15{\string\FU (#1){\equ(#1)}}
 \write16{ EQ #1 ==> \equ(#1) }}
\def\getichetta(#1){Fig. \verafigura
 \SIA g,#1,{\verafigura}
 \global\advance\numfig by 1
 \write15{\string\FU (#1){\graf(#1)}}
 \write16{ Fig. #1 ==> \graf(#1) }}
\def\retichetta(#1){\numpag=\pgn\SIA r,#1,{\verapagina}
 \write15{\string\FU (#1){\rif(#1)}}
 \write16{\rif(#1) ha simbolo  #1  }}
\def\etichettan(#1){(n\verocapitolo.\veranformula)
 \SIA e,#1,(n\verocapitolo.\veranformula)
 \global\advance\numnf by 1
\write16{\equ(#1) <= #1  }}
\fi

\newdimen\gwidth
\gdef\profonditastruttura{\dp\strutbox}
\def\senondefinito#1{\expandafter\ifx\csname#1\endcsname\relax}
\def\BOZZA{
\def\alato(##1){
 {\vtop to \profonditastruttura{\baselineskip
 \profonditastruttura\vss
 \rlap{\kern-\hsize\kern-1.2truecm{$\scriptstyle##1$}}}}}
\def\galato(##1){ \gwidth=\hsize \divide\gwidth by 2
 {\vtop to \profonditastruttura{\baselineskip
 \profonditastruttura\vss
 \rlap{\kern-\gwidth\kern-1.2truecm{$\scriptstyle##1$}}}}}
\def\verapagina{
{\romannumeral\number\numcap}.\number\numsec.\number\numpag}}

\def\alato(#1){}
\def\galato(#1){}
\def\veroparagrafo{\number\numsec}\def\veraformula{\number\numfor}
\def\veraappendice{\number\numapp}
\def\verapagina{\number\pageno}\def\veranformula{\number\numnf}
\def\verafigura{{\romannumeral\number\numcap}.\number\numfig}
\def\verocapitolo{\number\numcap}\def\veranformula{\number\numnf}
\def\Eqn(#1){\eqno{\etichettan(#1)\alato(#1)}}
\def\eqn(#1){\etichettan(#1)\alato(#1)}

\def\Eq(#1){\eqno{\etichetta(#1)\alato(#1)}}
\def\eq(#1){\etichetta(#1)\alato(#1)}
\def\Eqa(#1){\eqno{\etichettaa(#1)\alato(#1)}}
\def\eqa(#1){\etichettaa(#1)\alato(#1)}
\def\dgraf(#1){\getichetta(#1)\galato(#1)}
\def\drif(#1){\retichetta(#1)}

\def\eqv(#1){\senondefinito{fu#1}$\clubsuit$#1\else\csname fu#1\endcsname\fi}
\def\equ(#1){\senondefinito{e#1}\eqv(#1)\else\csname e#1\endcsname\fi}
\def\graf(#1){\senondefinito{g#1}\eqv(#1)\else\csname g#1\endcsname\fi}
\def\rif(#1){\senondefinito{r#1}\eqv(#1)\else\csname r#1\endcsname\fi}


\ifnum\tipoformule=1\let\Eq=\eqno\def\eq{}\let\Eqa=\eqno\def\eqa{}
\def\equ{}\fi



{\count255=\time\divide\count255 by 60 \xdef\hourmin{\number\count255}
	\multiply\count255 by-60\advance\count255 by\time
   \xdef\hourmin{\hourmin:\ifnum\count255<10 0\fi\the\count255}}

\def\oramin{\hourmin }

\def\data{\number\day/\ifcase\month\or gennaio \or febbraio \or marzo \or
aprile \or maggio \or giugno \or luglio \or agosto \or settembre
\or ottobre \or novembre \or dicembre \fi/\number\year;\ \oramin}

\setbox200\hbox{$\scriptscriptstyle \data $}

\newcount\pgn \pgn=1
\def\foglio{\number\numsec:\number\pgn
\global\advance\pgn by 1}

\def\foglioa{A\number\numsec:\number\pgn
\global\advance\pgn by 1}

\newskip\ttglue
\def\TIPIO{}
\def\TIPI{}
\def\TIPITOT{
\font\twelverm=cmr12
\font\twelvei=cmmi12
\font\twelvesy=cmsy10 scaled\magstep1
\font\twelveex=cmex10 scaled\magstep1
\font\twelveit=cmti12
\font\twelvett=cmtt12
\font\twelvebf=cmbx12
\font\twelvesl=cmsl12
\font\ninerm=cmr9
\font\ninesy=cmsy9
\font\eightrm=cmr8
\font\eighti=cmmi8
\font\eightsy=cmsy8
\font\eightbf=cmbx8
\font\eighttt=cmtt8
\font\eightsl=cmsl8
\font\eightit=cmti8
\font\sixrm=cmr6
\font\sixbf=cmbx6
\font\sixi=cmmi6
\font\sixsy=cmsy6
\font\twelvetruecmr=cmr10 scaled\magstep1
\font\twelvetruecmsy=cmsy10 scaled\magstep1
\font\tentruecmr=cmr10
\font\tentruecmsy=cmsy10
\font\eighttruecmr=cmr8
\font\eighttruecmsy=cmsy8
\font\seventruecmr=cmr7
\font\seventruecmsy=cmsy7
\font\sixtruecmr=cmr6
\font\sixtruecmsy=cmsy6
\font\fivetruecmr=cmr5
\font\fivetruecmsy=cmsy5
\textfont\truecmr=\tentruecmr
\scriptfont\truecmr=\seventruecmr
\scriptscriptfont\truecmr=\fivetruecmr
\textfont\truecmsy=\tentruecmsy
\scriptfont\truecmsy=\seventruecmsy
\scriptscriptfont\truecmr=\fivetruecmr
\scriptscriptfont\truecmsy=\fivetruecmsy
\def \eightpoint{\def\rm{\fam0\eightrm}
\textfont0=\eightrm \scriptfont0=\sixrm \scriptscriptfont0=\fiverm
\textfont1=\eighti \scriptfont1=\sixi   \scriptscriptfont1=\fivei
\textfont2=\eightsy \scriptfont2=\sixsy   \scriptscriptfont2=\fivesy
\textfont3=\tenex \scriptfont3=\tenex   \scriptscriptfont3=\tenex
\textfont\itfam=\eightit  \def\it{\fam\itfam\eightit}%
\textfont\slfam=\eightsl  \def\sl{\fam\slfam\eightsl}%
\textfont\ttfam=\eighttt  \def\tt{\fam\ttfam\eighttt}%
\textfont\bffam=\eightbf  \scriptfont\bffam=\sixbf
\scriptscriptfont\bffam=\fivebf  \def\bf{\fam\bffam\eightbf}%
\tt \ttglue=.5em plus.25em minus.15em
\setbox\strutbox=\hbox{\vrule height7pt depth2pt width0pt}%
\normalbaselineskip=9pt
\let\sc=\sixrm \normalbaselines\rm
\textfont\truecmr=\eighttruecmr
\scriptfont\truecmr=\sixtruecmr
\scriptscriptfont\truecmr=\fivetruecmr
\textfont\truecmsy=\eighttruecmsy
\scriptfont\truecmsy=\sixtruecmsy
}\def\nota{\noindent\eightpoint}}

\newfam\msbfam   
\newfam\truecmr  
\newfam\truecmsy 
\newskip\ttglue
\ifnum\tipi=0\TIPIO \else\ifnum\tipi=1 \TIPI\else \TIPITOT\fi\fi


\def\aps{{\it a posteriori}}
\let\0=\noindent\def\pagina{{\vfill\eject}}

\def\media#1{{\langle#1\rangle}}

\global\newcount\numpunt
\def\i#1{\def\9{#1}{\if\9.\global\numpunt=1\else\if\9,\global\numpunt=2
\else
\if\9;\global\numpunt=3\else\if\9:\global\numpunt=4\else
\if\9)\global\numpunt=5\else\if\9!\global\numpunt=6\else
\if\9?\global\numpunt=7\else\global\numpunt=8\fi\fi\fi\fi\fi\fi
\fi}\ifcase\numpunt\or{\accent18\char16.}\or{\accent18\char16,}\or
{\accent18\char16;}\or{\accent18\char16:}\or{\accent18\char16)}\or
{\accent18\char16!}\or{\accent18\char16?}\or{\accent18\char16\ \9}\else\fi}
\def\XWPR{{\it a priori}}
\def\ap#1{\def\9{#1}{\if\9.\global\numpunt=1\else\if\9,
\global\numpunt=2\else
\if\9;\global\numpunt=3\else\if\9:\global\numpunt=4\else
\if\9)\global\numpunt=5\else\if\9!\global\numpunt=6\else
\if\9?\global\numpunt=7\else\global\numpunt=8\fi\fi\fi\fi\fi\fi
\fi}\ifcase\numpunt\or{\XWPR.}\or{\XWPR,}\or
{\XWPR;}\or{\XWPR:}\or{\XWPR)}\or
{\XWPR!}\or{\XWPR?}\or{\XWPR\ \9}\else\fi}
\def\XWPSR{{\it a posteriori}}
\def\aps#1{\def\9{#1}{\if\9.\global\numpunt=1\else\if\9,
\global\numpunt=2\else
\if\9;\global\numpunt=3\else\if\9:\global\numpunt=4\else
\if\9)\global\numpunt=5\else\if\9!\global\numpunt=6\else
\if\9?\global\numpunt=7\else\global\numpunt=8\fi\fi\fi\fi\fi\fi
\fi}\ifcase\numpunt\or{\XWPSR.}\or{\XWPSR,}\or
{\XWPSR;}\or{\XWPSR:}\or{\XWPSR)}\or
{\XWPSR!}\or{\XWPSR?}\or{\XWPSR\ \9}\else\fi}

\def\ie{\hbox{\it i.e.\ }}\def\eg{\hbox{\it e.g.\ }}
\let\dpr=\partial\def\\{\hfill\break}

\def\*{\vglue0.3truecm}\let\0=\noindent
\let\==\equiv\let\txt=\textstyle
\let\io=\infty \def\V#1{\,\vec#1}   \def\Dpr{\V\dpr\,}
\let\ig=\int

\def\tende#1{\,\vtop{\ialign{##\crcr\rightarrowfill\crcr
              \noalign{\kern-1pt\nointerlineskip}
              \hskip3.pt${\scriptstyle #1}$\hskip3.pt\crcr}}\,}
\def\otto{\,{\kern-1.truept\leftarrow\kern-5.truept\to\kern-1.truept}\,}
\def\fra#1#2{{#1\over#2}}

\let\ciao=\bye
\def\fiat{{}}

\def\V#1{{\underline#1}}
\def\2{{1\over2}}
\def\EE{{\cal E}}
\def\CC{{\cal C}}\def\FF{{\cal F}}

\def\T#1{{#1_{\kern-3pt\lower7pt\hbox{$\widetilde{}$}}\kern3pt}}
\def\VV#1{{\underline #1}_{\kern-3pt
\lower7pt\hbox{$\widetilde{}$}}\kern3pt\,}
\def\W#1{#1_{\kern-3pt\lower7.5pt\hbox{$\widetilde{}$}}\kern2pt\,}

\def\NN{{\cal N}}

\def\lis{\overline}

\def\mbe{{\\*\hfill\hbox{\it
mbe\kern0.5truecm}}\vskip3.truept}

\def\TT{{\cal T}}
\def\1{{-1}}\def\qq{{\V q}}\def\pp{{\V p}}



\openin14=\jobname.aux \ifeof14 \relax \else \input \jobname.aux
\closein14 \fi
\openout15=\jobname.aux

\fiat

\vglue1cm
\centerline{\bf Topics on chaotic dynamics.}
\*
\centerline{Giovanni Gallavotti\footnote{${}^!$}{\nota
Dipartimento di
Fisica, $I^a$ Universit\`a di Roma, P.le Moro 2, 00185, Roma, Italia.
This paper is archived in {$mp\_arc@math.utexas.edu$}, \#94-333.}}
\*\*

\0{\it Abstract: Various kinematical quantities associated with the
statistical properties of dynamical systems are examined: statistics of
the motion, dynamical bases and Lyapunov exponents. Markov partitons for
chaotic systems, without any attempt at describing ``optimal results''. The
Ruelle principle is illustrated via its relation with the theory of
gases. An example of an application predicts the results of an
experiment along the lines of Evans, Cohen, Morriss' work on viscosity
fluctuations. A sequence of mathematically oriented problems discusses
the details of the main abstract ergodic theorems guiding to a proof of
Oseledec's theorem for the Lyapunov exponents and products of random
matrices.}
\*
\0{\it Keywords: \sl chaos, nonequilibrium ensembles, Markov
partitions, Ruelle principle, Lyapunov exponents, random matrices,
gaussian thermostats, ergodic theory, billiards, conductivity, gas.}
\*
\numsec=1\numfor=1
\0{\it\S1 Dynamical systems and their statistics.}
\*

A dynamical system $(\CC,S)$ will consist of a piecewise smooth compact
manifold and of a piecewise diffeomorphic map $S$ of $\CC$ into itself
with a piecewise diffeomorphic inverse $S^{-1}$.\footnote{${}^1$}{\nota
By piecewise smooth manifold I mean a manifold that can be regarded as
the union of a finite number of $C^\io$ compact manifolds, with any two
of them having in common only boundary points: the union of the
boundaries of such manifolds will be called the set of singularities of
$\CC$.  Similarly a map of $\CC$ into itself is said to be piecewise
smooth if $\CC$ can be regarded as the union of finitely many compact
$C^\io$ manifolds and $S$ is a diffeomorphism as a map of the interior
of each such manifold and its image.  A compact $C^\io$ manifold, of
dimension $d$, is a $C^\io$ manifold with boundary consisting of
finitely many compact $C^\io$ manifolds of dimension $<d$ with only
boundary points in common; the inductive definition is started by
declaring that a point is a $C^\io$ manifold of dimension $0$.} The set
$\NN$ of the singularities of either $S$ or of the manifold $\CC$ will
be called the set of the singularities of
$(\CC,S)$.\footnote{${}^2$}{\nota Sometimes one may wish to consider as
singular points some points where $S$ is in fact regular: in this case
such points will also be supposed to lye on piecewise smoooth
submanifolds of $\CC$ (of lower dimension) and will be included in the
set of singularities.  Furthermore, since the action of $S$ on the
singular points will not be studied, one may wish to require that
$S^{-1}$ be undefined or arbitrarily defined on the points $x\ni
S(\CC/\NN)$ so that in such points the property $S(S^{-1}x)=x$ may even
fail.  And likewise one may avoid defining $S$ on $\NN$.  I will not try
to be so general: see [P].}

The reader disturbed by the above generality can simply think that all
I am discussing is the case of a $C^\io$ diffeomorphism of a $C^\io$
compact manifold without boundary. The concession to the generality is
due to the fact that some of the most important dynamical systems
really show physically significant singularities, like the billiards
or the hard sphere gases.

\*
{\it Example 1:} A Hamiltonian system with $l$ degrees of freedom is
observed at the instants in which a certain event, {\it the timing
event}, happens.  This is the event in which the point representing the
system passes through a predefined surface in phase space (\eg one
particle of the system passes through an ideal wall in physical space
or the distance between two particles takes some prefixed value $r$).
If $\CC$ is the family of timing events with a prefixed energy $H_0$,
and $S$ is the transformation mapping one event in $\CC$ into the
following, then $(\CC,S)$ is a dynamical system. The probability
distribution $\m_L$ defined by assigning to a (measurable) set of events
$E\subset\CC$ the Liouville measure of the set of phase space points on
the energy surface experiencing the first timing event in $E$, is an
invariant measure (\ie $\m_L(E)=\m_L(S^{-1}E)$). The $\m_L$ will still
be called the Liouville measure.

{\it Example 2:} consider a box, $[-\fra12L,\fra12L]^2$, with side $L$
and periodic boundary conditions (\ie op\-po\-si\-te sides identified)
and with a few circular regions $C_i$, "obstacle of radius" $r_i$, in
it.  A particle moves freely among them and collides elastically, at
velocity $v$.  The space $\CC$ consists of the {\it collisions},
parameterized by the point $\a$ on the obstacle where the collision
takes place and by the angle $\f$ formed between the incoming velocity
and the outer normal to the collision point: hence
$\f\in[-\fra\p2,\fra\p2]$.  We see that $\CC$ consists in the union of
$\dpr C_i\times[-\fra\p2,\fra\p2]$.  A point in $\CC$ is determined by
$2$ coordinates $(\a,\f)$, with $\a\in [0,2\p]$ being the angular
position of the collision point and a label $i$ denoting the obstacle on
which the collision takes place.  The obstacles will be supposed so
distributed so that no collisionless trajectory is possible.  The map
$S$ is smooth everywhere except on the boundary of $\CC$ (\ie
$\f=\pm\p$) and on the collisions $x=(i,\a,\f)$ such that $Sx$ is a
tangent collision.  This is the {\it billiard system}.  A natural metric
on $\CC$ is $d\a^2+d\f^2$.  The system being hamiltonian it is easy to
find the Liouville distribution: in the coordinates $(\a,\f)$ it is
given by $\G r_id\a\,\cos\f\,d\f$ where $(\a,\f)$ represent collisions
with the obstacle $i$ and $\G$ is defined so that $\m_L(\CC)=1$.

{\it Example 3:} many particles, say $N$, in a periodic box as in the
example 2.  The particles interact with a radial pair potential $v$
(with range smaller than $L$) and with an external potential
represented by the hard core due to the obstacles.  The phase space is
$\CC^N$ and the equations of motion are:

$$ \V{{\dot q}}_j=\fra1m\V p_j,\quad\V{{\dot p}}_j=\V F_j+E\V i-\a\V
p_j\Eq(1.1)$$

\0where $\V i$ is the $1$--axis unit vector, $E$ is an external
constant field, $\V F_j$ is the force generated by the total potential
$\F$ on the $j$--th particle; the obstacles are taken into account by
the elastic reflection rule and $\a$ is so defined that the energy
$H=\sum_j\fra1{2m}\V p_j^2+\F$ is a constant of motion ({\it gaussian
thermostat}): \ie $\a=E\V i\cdot\sum_j\V p_j/\sum_j \V p_j^2/2m$, as a
simple calculation proves; $m$ is the particles mass.  The phase space
$\CC$ consists of the energy $H_0$ configurations in which one
particle (any one) is colliding with some obstacle.  The time
evolution $S$ maps one such configuration into the following
configuration of the same type ("next collision configuration").  The
phase space thus defined is a $4N-2$ dimensional subspace of the total
$4N$ dimensional phase space; and a simple calculation of the
divergence of the r.h.s.  of \equ(1.1) shows us that the phase space
volume changes at a rate $(2N-1)\a$ in the total $4N$ dimensional
phase space $\FF$.  This example has a rather involved set of
singularities: nevertheless it is piecewise smooth and it defines a
dynamical system in the above sense.  Physically it is a model for
electric conductivity ({\it Lorentz gas conductor}).

{\it Example 4:} The map of the torus $\CC=[0,2\p]^2$ defined by:
$S(\f_1,\f_2)=(\f_1+\f_2,\f_1+2\f_2)\,{\rm mod}\,2\p$. One also
writes:

$$S\pmatrix{\f_1\cr\f_2\cr}=\pmatrix{1&1\cr1&2\cr}\pmatrix{\f_1\cr\f_2
\cr}\,{\rm mod}\,2\p\Eq(1.2)$$

\0Then $S$ preserves the distribution $\m_0(d\V\f)=d\V\f/(2\p)^2$.  This
system has {\it no singularities}.  \*

Given a dynamical system $(\CC,S)$ we shall consider initial data chosen
randomly with respect to a given probability distribution $\m_0$. It is
customary to consider the "Liouville distribution": the latter means a
probability distribution with a positive density with respect to the
volume on the phase space, as one often calls the space $\CC$.
Sometimes it could literally coincide with the Liouville measure of
analytical mechanics; \eg when the system $(\CC,S)$ is obtained by
timing the evolution of a hamiltonian system: this is the case in
examples 1,2.

The {\it qualitative theory} of $(\CC,S)$ studies the asymptotic
properties of the motions following initial data chosen randomly with
distribution $\m_0$, which we call ``$\m_0$--random data''.

More generally one could investigate the properties enjoyed by motions
with initial data chosen randomly with respect to other probability
distributions $\m$ not absolutely continuous with respect to the
volume measure. Sometimes this is regarded as a less interesting
question on the grounds that the Liouville distribution is the {\it
natural one} to use in selecting initial data. This is a
preconception; in fact it is well known that it is equally easy (or,
rather, difficult) to produce random initial data with a distribution
singular with respect to the uniform distribution: therefore such data
are equally interesting; or at least one has to find better reasons to
regard them less interesting. This being not the place to undertake a
learned philosophical discussion on the {\it preminence} of the
Liouville measure $\m_0$, I shall concentrate on the most
studied question of which is the asymptotic behaviour of motions with
initial data chosen randomly with distribution $\m_0$. Many of the
properties that will emerge will be relevant also for the other random
choices with respect to less appreciated distributions.

The key notion is that of {\it statistics}: \* {\bf 1 Definition:}
{\it Given a dynamical system $(\CC,S)$ and a probability distribution
$\m$ attributing probability $1$ to the set of points $x$ which
never in their evolution fall on the singularity set $\NN$ (\ie
$\m(\cap_{k=0}^\io S^{-k}(\CC/\NN))=1$), we say that $\m_0$ has a
{\sl statistics} with respect to the evolution $S$ if for any
continuous function $F$ on $\CC$, called {\sl observable}, the time
average of $F$ on the motion generated by $\m$--almost all points
$x\in\CC$ exists, and has the form:

$$\lim_{T\to\io}
\fra1T\sum_{k=0}^{T-1} F(S^kx)=\ig_\CC F(y)\lis\m(dy)\Eq(1.3)$$

\0where $\lis\m$ is a suitable probability distribution on $\CC$,
called the {\sl statistics of $\m$} in the dynamical system $(\CC,S)$.
If $\m=\m_0$ is the Liouville distribution then $\lis\m$ will be called
simply the statistics of $(\CC,S)$ and we shall say that "$(\CC,S)$ has a
statistics" without reference to $\m_0$.}\footnote{${}^3$}{\nota
Adhering, only for lack of time, to the prejudice that the volume
measure is somewhat privileged.} \*

\0{\it Remarks: }
\*
\0(i) Usually the map $S$ is such that $S^{-k}\NN$ is contained in a lower
dimensional compact submanifold of $\CC$: therefore almost all points
(randomly chosen with respect to the Liouville measure) will never fall
on $\NN$ and it will make sense to ask if the Liouville measure has a
statistics.

\0(ii) Note that $\lis\m$ {\it does depend on $\m_0$}. It makes no sense to
talk about the statistics of a dynamical system without specifying the
distribution with which the initial data are randomly selected.

\0(iii) If $(\CC,S)$ has a statistics (\ie if the Liouville distribution
has a statistics) then one says also that $(\CC,S)$ has a {\it unique
attractor}.  It might happen, and in some interesting cases it does
happen, that the space $\CC$ can be represented as a union of several
open sets $U_i$ and of some $0$ $\m_0$--measure sets $N_k$: for almost
all points in $U_i$ the average of $F$ is given by a formula like
\equ(1.3) but with some $\lis\m_i$ replacing $\lis\m$.  In this case one
would say that $(\CC,S)$ has several attractors each with its own
statistics.  In the present analysis I shall not consider such cases:
often they can be reduced to cases with unique statistics simply by
redefining the phase space to be $\lis U_i$ rather than $\CC$; although
this is not the most general case.

A further general notion is that of {\it attractor}:
\*
{\bf 2 Definition:} {\it The set $A\subset \CC$ is an "attractor"
for  $\m$--random data if $\m$ has a statistics
in the dynamical system $(\CC,S)$ and (1): $SA=A$ ("$A$ is invariant"),
(2): $\lis\m(A)=1$ ($A$ has "probability $1$ with respect to the
statistics),
(3): the fractal (Hausdorff)
dimension of $A$ is minimal among the $A$'s with the
properties (1),(2).}

\*
\0{\it Remarks:}
\*
\0It is not convenient to require that the attractor be
a closed set (as sometimes done).  In many cases the attractors $A$ in
the above sense are dense in $\CC$ but have a fractal dimension strictly
less than that of $\CC$.  So the above notion is subtler than the ones
that imply or require that an attractor be closed.  It also stresses
that an attractor will usually have a non trivial fractal dimension. On
the other hand the attractor in this sense is not unique (\eg one can
usually remove from it countably many points, except in special cases
like when $A$ is a fixed point or a periodic orbit, or even larger sets
with zero Hausdorff dimension and zero $\lis\m$ measure). \*

With the above notions in mind we can proceed to define some of the main
qualitative features of the motions following data chosen randomly with
respect to a distribution $\m_0$.

\*
\0{\it\S2 Dynamical bases and Lyapunov exponents.} \*
\numsec=2\numfor=1

A given motion analysis can starts with the attempt at
understanding the behaviour of nearby trajectories: this means
understanding the linearization of the motions taking place near
it.

For this purpose the notion of regular point is necessary:\*

{\bf 3 Definition: \it A point $x$ is {\sl regular} if the map $S$ is
regular in the vicinity of $S^kx$ for all $k\ge0$.  Recalling that $S$
is always supposed to be piecewise regular this simply means that
$S^kx$ does not fall into the set $\NN$ of singularities of $S$: \ie
$x\not\in \cup_{k=0}^\io S^{-k}\NN$ or $x\in\cap_{k=0}^\io
S^{-k}(\CC/\NN)$.}\*

Let $S^kx$ be a trajectory starting at a regular point $x$,
$k=0,1,\ldots$: the trajectory $S^k(x+dx)$ with $dx$ infinitesimal
will stay close forever to that of $x$ departing from it by an
infinitesimal amount simply given by $\dpr S^k(x) dx$, where $\dpr
S^k(x)$ is the jacobian matrix of $S^k$ evaluated at the point $x$.
It is a $d\times d$ matrix if $d$ is the dimension of $\CC$; by the
chain rule $\dpr S^k(x)$ is :

$$T_k(x)\=\dpr S^k(x)=\dpr S(S^{k-1}x)\cdot\ldots\cdot\dpr S(x)\=
\prod_{j=1}^k\dpr S(S^{k-j}x)\Eq(2.1)$$
where the factors appear ordered from left to right as $j$ increases.
\*

{\it I shall assume that for some $\bar C,\bar \e>0$ it is $|\det\dpr
S(x)|\ge\bar\e>0$ and $|\dpr S(x)|<\bar C$}, for all regular points
$x$. This assumption could be greatly weakened, see the appended
problems and [P], but it is convenient continuing the discussion
without having to worry about such matter.

Therefore the square of the {\it stretching} of $dx$ will be described
by the ratio:
$$\fra{(\dpr S^k dx,\dpr S^k dx)}{(dx,dx)}\Eq(2.2)$$
where the scalar products are evaluated in the metric defined on $\CC$.
If $G(x)$ is the metric tensor in $x$ and if the {\it dilatation}
matrix $(M_k)_{ij}$
is defined by setting: $M_k=(\dpr S^k(x))^*\dpr S^k(x)$, we see that in
the local coordinates the matrix is expressed by:
$(M_k)_{ij}=(\dpr S^k(x))^TG(S^kx)(\dpr S^k(x))_{ij}$.  So that
\equ(2.2) can be explicitly written as:
$$\fra{\sum_{ij} (M_k)_{ij}dx_i\,dx_j}{\sum_{ij} G(x)_{ij}dx_idx_j}
\Eq(2.3)$$
Suppose that the symmetric matrix $M_k^{1/2k}$ has a limit $D_+(x)$, as
$k\to\io$.  Then tere is $\tilde c$ such that $\tilde c>|\det D_+(x)|\ge
\bar\e$ by the assumption, and we can define $l_1(x)\ge l_2(x)\ge\ldots
l_d(x)>0$ to be its ordered eigenvalues, counted according to
multiplicity.  Let $\V v_1(x),\ldots,\V v_d(x)$ be the corresponding
eigenvectors, which can be taken orthonormal with respect to the "naive"
scalar product $(a,b)=\sum_i a_i b_i$ (which is the scalar product in
which $M_k$ and $D_+$ are symmetric).  Then $M_k \V v_j(x)$ grows as
$l_j(x)^k$, in the sense that $\fra1k \log|M_k \V v_j(x)|\tende{k\to\io}
l_j(x)$, as an elementary estimate shows.

We also call $\lis l_1(x)>\lis l_2(x)>\ldots> \lis l_{s(x)}(x)$,
respectively $\lis \l_1(x)>\lis \l_2(x)>\ldots> \lis \l_{s(x)}(x)$, the
{\it distinct eigenvalues} (respectively their logarithms) of $D_+(x)$
and $n_1(x),\ldots,n_{s(x)}(x)$ their multiplicities, \ie the dimensions
of the eigenspaces $U_1(x),\ldots,U_{s(x)}(x)$ spanned by the
eigenvectors with the same eigenvalue.  The $l_j(x),\l_j(x)$ can be
called the spectral or scaling coefficients at $x$ and, respectively,
the spectral or scaling exponents at $x$.

Then it is clear that the vector space $R^d$ can be regarded as
containing the planes $V_1(x)\=R^d\supset V_2(x)\supset V_3(x) \supset
V_{s(x)}(x)$ where $V_j(x)$ is the plane spanned by the eigenvectors
of $D_+(x)$ with eigenvalue $\le \lis l_j(x)$: the dimensions of
$V_j(x)$ are $d_j=n_j(x)+\ldots+n_{s(x)}(x)$ and the planes $V_j(x)$
are $V_j(x)=U_j(x)\oplus \ldots\oplus U_{s(x)}(x)$.

It is very important to realize that the vectors $\V v_j$ {\it do
depend on the choice of coordinates (and of the metric} $G$): by
changing the metric or the coordinates the vectors $\V v_j$ may, in
general, change.  However the eigenvalues $\lis l_j(x)$, and their
multiplicities $n_j(x)$ cannot change, and
{\it also the sequence of decreasing subspaces $V_j(x)$ cannot
change}, as one can see by noting that the scaling
propeties of the vectors in such spaces are not metric poperties, but
have an intrinsic geometric meaning. They can be {\it characterized}
by the property that $\fra1k \log|\dpr
S^k(x)u|\tende{k\to+\io}\lis\l_j(x)$ if $u\in V_j(x)/V_{j-1}(x)$.

One could try to define the planes with given contraction rate $\lis
\l_j$ to be just $U_j(x)$.  But such spaces would not have an intrinsic
dynamical meaning: because one could "tilt" $U_j(x)$ sligthly and
still keep the property that all the vectors contract at the rate
$\lis\l_j$ (provided the tilting does not generate components along the
$U_{j+p}(x)$ with $p>0$, of course).

Likewise one cannot assign a special meaning to the planes
$U_1(x)\oplus\ldots \oplus U_{j}(x)$: this would in fact generate a
system of planes such that if $u$ is a vector in the $j$-th plane and
not in the $(j-1)$-th then the exponent of dilatation will be
$\lis\l_j$; {\it but such system is not uniquely determined}, for the
same reasons discussed in the previous paragraph. On the other hand
the system $V_j(x)$ has the property of being uniquely determined by
the action of $S$, and it defines a system of planes along which the
scaling size becomes weaker.  \*

{\bf 4 Definition: \it Given the dynamical system $(\CC,S)$ a point
$x\in\CC$ admits a "{\sl system of scaling (or contracting) planes}"
for the forward motion, or for $S$, if:\\ (1) $x$ is a regular
point.\\ (2) there exist numbers $\lis\l_j(x)$ and positive integers
$s(x)$ and $n_j(x)$, with $j=1,\ldots,s(x)$, with the properties:\\
(3) the space $R^d$ of the infinitesimal vectors out of $x$ can be
regarded as containing a sequence of $s(x)$ subspaces
$V_1(x)\=R^d\supset V_2(x)\supset\ldots \supset V_{s(x)}(x)\supset
V_{s(x)+1}\=\V0$ with $V_j(x)$ having dimension
$n_{s(x)}+\ldots+n_j(x)$ and\\ (4) the following limits hold:

$$\lim_{k\to+\io} \fra1k \log |\dpr S^k(x) v| =\lis\l_j(x)\qquad {\rm
if}\ v\in V_j(x)/V_{j+1}(x)\Eq(2.4)$$

\0Then one says that the numbers $\lis\l_j(x)$ are the {\sl scaling
exponents} of $S$ at $x$ in the forward direction, their exponentials
$l_j(x)=e^{\lis\l_j(x)}$ are the {\sl scaling coefficients}, the number
$n_j(x)$ is the multiplicity of the $j$-th coefficient.  Sometimes the
coefficients are repeated according to the multiplicity: in this case
their number is, of course, exactly the phase space dimension. The
scaling exponents (coefficients) for $S$ are often called the {\sl
forward Lyapunov} exponents (coefficients)}.

\*
\0{\it Remarks:}\*

\0(i) There is no reason why a point should admit a system of
scaling planes.\\
(ii) A sufficient condition for the existence of contracting planes is
that the limit of the sequence of matrices
$(\dpr S^k(x)^*\dpr S^k(x))^{1/2k}$ exists and is a positive
matrix $D$, see the discussion preceding the definition 4,
\\
(iii) If $\lis\l_1(x)>0$, $\lis\l_{s(x)}<0$ and $\lis\l_j(x)\ne0$ for
all $j$'s one says that $x$ is a {\it hyperbolic point}.  In this case
if $r_-$ is such that $\lis\l_j>0$ for $j< r_-$ and $\lis\l_j<0$ for
$j\ge r_-$ the plane $V^s=V_{r_-}$ will be called the {\it contraction},
or {\it stable, plane}. The motion of a hyperbolic point is very
unstable in, essentially, all possible senses.
\\
(iv) If a point $x$ admits a system of scaling planes so do all the
points on the trajectory generated by $x$.  Such points do have the same
coefficients (multiplicities included) and the spaces $V_j(x)$ and
$V_j(Sx)$ are related by: $V_j(Sx)=\dpr S(x) V_j(x)$, for all $j$'s. The
spaces $U_j(x)$ instead {\it do not have, in general, any covariance
property} because they are not intrinsically defined by the dynamics
(see above).
\\
(v) If $x$ admits a system of scaling planes
then it may not admit such a system in the dynamical system
$(\CC,S^{-1})$; and even if it does there is no reason why there should
be any relation between the two systems or the relative exponents,
c.f.r. problem (32).
\\
(vi) Note, once more, that the subspaces in which the contraction
exponent is $< \l$ for some real $\l$, intermediate between the
exponents values, are well defined (and coincide with $V_j(x)$ if
$\lis\l_{j-1}>\l>\lis\l_{j}$). But the ones in which the rate is $>\l$
are {\it not} well defined and this explains the impossibility of
associating expanding planes to the forward motion.
\\
(vii) Note that the scaling planes for $S$ at a point, when existent,
concern the {\it forward} motion: \ie they are properties of the
trajectory $S^k x,\,k\ge0$.  Likewise the scaling planes for $S^{-1}$
concern only the {\it backward} motion.

\*
In view of the above remarks it is important to try to establish some
general results about the existence of systems of scaling planes.

A simple case concerns the regular hyperbolic fixed points $x$: the
latter always admit a system of scaling planes for the forward and
backward motion, \ie for $S$ and for $S^{-1}$.  The linearization of
the map $S$ around $x$, \ie the matrix $T=\dpr S(x)$ defines the
$V^s_x$ plane spanned by the spectral planes of the eigenvalues of $T$
with modulus $<1$ and the subspace $V^u_x$ spanned by the planes
relative to the eigenvalues of $T$ with modulus $>1$.  The two subspaces
$V^s_x,V^u_x$ can be continued into two small regular connected
manifolds tangent to them in $x$, $\D^s_x,\D^u_x$, such that if $y\in
\D^s_x$ then $|S^{ k}y-S^{ k}x|\tende{k\to\io}0$ bounded
proportionally to $ l^{k}_-$, if $ l_-$ denote the absolute values of
the eigenvalue of $T$ closest to the unit circle and $<1$. Or,
similarly, if $y\in
\D^u_x$ or $>1$ then $|S^{- k}y-S^{- k}x|\tende{k\to\io}0$ bounded
proportionally to $ l^{-k}_+$, if $ l_+$ denotes the absolute value of
the eigenvalue of $T$ closest to the unit circle and $>1$.

In this case the forward scaling coefficients coincide with the
absolute values of the eigenvalues of $D$, called in stability theory
the {\it Lyapunov coefficients} of the fixed point.  They can also be
defined as the eigenvalues of the matrix $D$ (whose existence is not
completely trivial, see problems):
$\lim_{n\to\io}((T^*)^nT^n)^{1/2n}=D$.  The backward scaling
coefficients can likewise be identified with the eigenvalues of the
matrix $\lim_{n\to\io}((T^{-k})^*T^{-k})^{1/2k}=D_-$ and, therefore,
they are the reciprocals of the forward coefficients. The contracting
plane $V^u_x$ for $S^{-1}$ is, also, spanned by the eigenplanes of
$D_-$ corresponding to the eigenvalues $<1$.

It is convenient to set up the following definition for future use:
\*
{\bf 5 Definition:} {\it If $x$ is a fixed point the set $\cup_{k=0}^\io
S^k\D_x^u\=W^u_x$ will be called the "global unstable manifold", while
the set $\cup_{k=0}^\io S^{-k}\D_x^s\=W^s_x$ will be called the global
stable manifolds of $x$.}
\*

The  two manifolds are locally regular around all their points that do
not fall on the singularities of $S^k$ or $S^{-k}$, respectively, for
some $k\ge0$. But they may be even disconnected, in general.

The subspace tangent to $\D_x^s$ at $x$ coincides with the plane
generated by the vectors associated with the Lyapunov numbers less
than $1$ which, as said above, is intrinsically associated with $S$.
It consists of the infinitesimal vectors that contract at exponential
rate under the action of $S$.

One cannot identify "similarly" the subspace tangent to $\D_x^u$, at
$x$, as the plane generated by the infinitesimal vectors expanding at
exponential rate under $S$ because, as already pointed out, such vectors
are not unambiguously defined (given a vector expanding at exponential
rate one can add to it a vector that expands at a lower rate, getting a
vector expanding at the same rate). However one can identify $\D^u_x$ in
the same way as $\D^s_x$ by replacing $S$ with $S^{-1}$.

The extension of the analysis to the case of a periodic point is very
easy and well known and it will be skipped.  In the case $x$ is not
fixed, nor periodic, the discussion above leads to a natural extension
expressed by the definition: \*

{\bf 6 Definition:} {\it A point is said to admit a dynamical base
$(Z_1,\ldots,Z_s)$, consisting of $s$ mutually transversal planes, if:\\
(1) it is regular for $S$ and $S^{-1}$ and it admits, both in the
forward and in the backward motions, scaling systems of planes with
opposite Lyapunov exponents.\\ (2) no Lyapunov exponent vanishes.\\ (3)
the following limit relation holds as $k\to+\io$ as well as $k\to-\io$:

$$\lim_{k\to\pm\io}\fra1k \log |\dpr S^k u|=\lis\l_j(x)\qquad{\rm for}\
u\in Z_j\Eq(2.5)$$

\0In this case the planes $V_j=Z_s\oplus\ldots\oplus Z_{j}$ will be
called the system of {\sl contraction planes} for $S$ and the $\tilde
V_j=Z_1\oplus\ldots \oplus Z_j$ will be the system of {\sl expanding
planes} for $S$.  If $s_-$ is such that $\lis\l_j<0$ for $j\ge s_-$
and $\lis\l_{j}>0$ for $j< s_-$ the planes $V^s=V_{s_-}$ and
$V^u=\tilde V_{s_--1}$ will be called the {\sl stable and unstable
planes} of $S$ at $x$.} \*

\0{\it Remarks:} \*
\0(a) The corresponding notions for the backward motion (\ie for
$S^{-1}$) are trivially related to those for $S$.\\
\0(b) If $u\in V_j(x)/V_{j+1}(x)$ then \equ(2.4) holds.\\
\0(c) The points $S^kx$ also admit a dynamical base and $Z_j(S^kx)=\dpr
S^k(x) Z_j(x)$ for all (signed) integers $k$.
\*

Finally one more definition is useful to simplify the language:
\*
{\bf 7 Definition:} {\it A point $x$ is called {\sl normal} if it admits a
dynamical base.}
\*

Hence a normal point is a generalization of a periodic point.  And it is
remarkable that such points do exist and in fact abund, in some sense.
The first classical result concerns invariant ergodic distributions $\m$
on $\CC$: here invariance means that $\m(\cap_{-\io}^\io S^k\NN))=0$ and
$\m(S^{-1}E)=\m(E)$ for all (Borel)\footnote{${}^*$}{\nota This is in
parenthesis because all the sets I will mention are Borel sets or differ
from a Borel set by a set of measure $0$ with respect to the measure
that is being considered and which I also call for brevity Borel sets if
no ambiguity arises. In fact, I cannot even conceive the other sets.}
sets $E$ and ergodicity means that there are no non trivial constants of
motion which are not $\m$--almost surely constant.

\*
{\bf I Theorem:} {\it Let $\m$ be an invariant ergodic distribution for
$(\CC,S)$: then $\m$ almost all points are normal for $S$.  Furthermore
the Lyapunov exponents and their multiplicities are almost everywhere
constant.\\ If no contraction or expansion coefficient has value $1$ and
if $V^s_x$ and $V^u_x$ are the contracting and expanding planes, then
under further "weak" regularity assumptions, there exist ({\sl
integrability property}) manifolds $W^u_x$ and $W^s_x$ tangent in $x$ to
$V^u_x$ and $V^s_x$, smooth near $x$, such that:
$$\eqalign{
|S^kx-S^ky|&\tende{k\to\io}0,\quad\hbox{for}\ y\in W^s_x\cr
|S^{-k}x-S^{-k}y|&\tende{k\to\io}0,\quad\hbox{for}\ y\in
W^u_x\cr}\Eq(2.6)$$
and the approach to $0$ takes place exponentially fast bounded above
proportionally to $e^{-\l k}$, where $\l>0$ is a suitable constant
(such that no Lyapunov
exponent is in the interval $[-\l,\l]$).}
\*
A discussion of the "further regularity assumptions" would lead us away
from the themes chosen here for discussion: I just mention that they are
assumptions on the speed at which a regular point $x$ can approach, in
its evolution, the singularities, [P]. The idea is that such speed
should be slower than any exponential (or at least slower than the speed
corresponding to a bound $\l$ on the contraction and expansion rate).

The above first statement is the {\it Oseledec theorem}: a "guided"
proof to it is described in the problems, the statements concerning the
integrability property are part of Pesin's theory, [P]. The theorem
shows that if the data are picked up randomly with respect to an ergodic
distribution (any one!) then they provide (non constructive) examples of
points with dynamical bases. It also shows that the scaling properties
of $S$ and those of $S^{-1}$ are intimately related and the two maps
show the "same" properties (or, rather, properties that are trivially
related), provided the data are randomly distributed with an {\it
invariant} ergodic distribution giving zero measure to data visiting, in
their evolutions, the singularities.  "Nothing about data randomly
chosen with an egodic invariant distribution can be learnt by running
the motion backwards, that cannot be learnt by running it forward". In
other words one can say that the motion is "reversible" {\it on such
data}.\footnote{${}^4$}{\nota Remember, however, that such data are very
special.}

The case in which there are $0$ contraction or expansion exponents is
more involved and it will not be discussed. It certainly arises when
the system has non trivial smooth constants of motion, each of which
generates a vanishing Lyapunov exponent. But such cases can be easily
eliminated because one usually can restrict the phase space to the
data for which the constants of motion have a fixed value. When zero
Lyapunov exponents are not related to smooth constants of motion,
however, a genuinely more complicated situation arises: I shall not
deal with it here, besides saying that some of the above kinematical
properties {\it do not depend on the assumption of the absence of $0$
Lyapunov exponents}. In particular the first sentence in the theorem I
above {\it does not require} that the exponents be non zero.

Also the ergodicity, in some sense, is not really necessary and it has
been introduced only to simplify the exposition.  The case in which
ergodicity is missing can be often reduced to the ergodic
case by using the "ergodic decomposition theorem": a theorem and a
discussion that I avoid here.

A more interesting question is what can be said when the data are picked
up with a distribution that is {\it not} invariant.  The case of non
invariant distribution of the initial data will be analyzed, under
suitable further assumptions, in the next section.  Here I add only a
few remarks on a frequently misrepresented procedure.

Note that the Lyapunov exponents for $S$ and those of $S^{-1}$ are
opposite {\it if the data are chosen randomly with respect to an
invariant distribution.} However usually one selects the data with
respect to a {\it non invariant distribution}: in this case the above
theorem does not say much. {\it In particular one should expect that
the asymptotic properties of the motions in the future and in the past
are different and described by different statistics}. For instance in
cases in which there is a {\it time reversal symmetry} one will find
that the Lyapunov exponents for the motions towards the past and the
future are identical (rather than opposite). This will not contradict
the fact that if the data are chosen randomly with respect to the {\it
future} statistics then the Lyapunov exponents for the motion towards
the past would be the {\it opposite} of the ones for the motion
towards the future!

How comes that often a method to measure the minimum Lyapunov exponent
is {\it said } to be: "just measure the maximum Lyapunov exponent for
the backward motion and change sign"? In most cases this would seem
wrong: for instance when the system is time reversible
{\it because} one would, instead, get again the maximum Lyapunov
exponent.

The "paradox" is understood if one examines what is really meant by the
above statement, \ie what is the actual measurement performed.  One
finds that the measurement consists (schematically) in picking at
random with a distribution $\m_0$ (non invariant) a point $x$,
following its trajectory for a long time $T$, togheter with the
trajectory of a nearby point $y$ until it reaches a point $S^Tx$.  The
rate of separation of the two points gives\footnote{${}^5$}{\nota
Unless the two points are very special, \eg if $y$ lies on the lower
dimensional contracting manifold $W^s_x$.} a measurement of the
maximum Lyapunov coefficient.  The trajectory of $x$ is memorized and
that of $y$ "thrown away".  One then starts from $S^T x$ and from a
nearby point $y'$, and runs the two motion backwards: of course the
motion of $S^Tx$ is already known and one computes that of $y'$.  The
motions of both points  $y,y'$ is computed as a perturbation of the
motion of $x$ or of $S^Tx$ (forward and backwards respectively) thus
it is very convenient to use twice the same trajectory as this saves
considerable time.

But this is not merely a matter of convenience: we see that $S^Tx$ is
{\it no longer a random point with the initial distribution}. It is,
rather, approximately a random point with respect to the statistics
$\lis \m$ of the distribution $\m_0$.  Therefore it will show in the
backward motion Lyapunov exponents opposite to the ones exhibited in the
forward motion (which have the {\it same} value if the initial data
are chosen with respect to $\m_0$ or $\lis\m$), by the above theorem.
Hence one gets the wanted result: in the {\it improper sense just
discussed} the maximum exponent in the backward motion is the opposite
of the minimum in the forward motion. This property is very reminiscent
of the build up of correlations that appears in the theory of the
molecular chaos in the Boltzmann equation, see [CB].

\*
\0{\it\S3 Chaotic motions.}
\numsec=3\numfor=1\def\qq{{\V q}}
\*
The definition that I shall use for a chaotic system is:

\*
{\bf 8 Definition:} {\it A dynamical system $(\CC,S)$ will be called
"chaotic" if:\\ (1) There exists a periodic hyperbolic point $O$ such
that the global stable and unstable manifolds of $O$ consist of regular
points for $S^{\pm1}$, and are smooth, connected and dense on $\CC$,
({\sl instability axiom}).\\ (2) The restrictions of $S^k$ to $W^u_O$,
and of $S^{-k}$ to $W^s_O$, are uniformly expansive for $k\ge0$ and
uniformly contractive for $k\le0$, ({\sl expansivity axiom}).\\
\0(3) If $x_n,x'_n$ are points on $W^u_O$ and
$|x_n-x'_n|\tende{n\to\io}0$ then the tangent planes to $W^u_O$ at such
points become parallel uniformly at speed $|x_n-x_n'|^\b$ for some
$\b>0$, togheter with the matrix that linearizes the evolution $S$ on
$W^u_O$: a similar property holds for $W^s_O$; ({\sl continuity
axiom}).\\
\0(4) The manifolds $W^s_O,W^u_O$ form, everywhere they cross, an angle
uniformly bounded away from $0$, ({\sl transversality axiom}).\\
(5) Given any open sets $G, \G_1,\ldots,\G_n$ there is a $k$ such that
$S^h G$ has intersection with all the $\G_j$ for $h\ge l$ ({\sl
topological mixing axiom}).}
\*
\0{\it Remarks:}
\*

\0(i) To understand the role of the various conditions in the above
definition note that from the hyperbolicity of $O$, and from the
definitions of the previous section, it follows that the tangent planes
to the manifolds $W^u_O$ and $W^s_O$ are the expansive and contractive
scaling manifolds of $O$.  However it does not follow from this only
fact that the action of $S$ on $W_O^u$, nor that the action of $S^{-1}$
on $W^s_O$, are expansive. The latter property, as well as any statement
on the Lyapunov exponents on $W^u_O$, concern properties of the {\it
future} evolution on $W_O^u$, while the knowledge that $W^u_O$ is the
unstable manifold of the fixed point gives us only information about the
motion towards the past.  Hence it cannot follow that $(\dpr
S^k(x)^*_u\dpr S^k(x)_u)^{1/2k}$, where $\dpr S(x)_u$ denotes the
jacobian of the transformation $S$ regarded as a map of $W^u_O$, has a
limit as $k\to\io$.  The role of (2) is to require such properties, or
at least the part of them that we need, explicitly (as well as the
corresponding ones for $W^c_O$). Its quantitative meaning is expressed
as: there exist $D_0,\l>0$ such that for all $x,y\in W^u_O$ and $k\ge0$
the distance $d_u(S^{-k}x,S^{-k}y)$ between $S^{-k}x,S^{-k}y$, {\it
measured along $W^u_O$}, is:

$$d_u(S^{-k}x,S^{-k}y)\le D_0 e^{-\l k}
d_u(x,y)\Eq(3.1)$$

\0and the corresponding properties are required to hold for $W^s_O$
for $k\le0$.

\0(ii) Property (3) on the jacobians means that there exists $D_1,\b>0$
such that the "unit vector orthogonal to $W^u_O$ at $x$", which is in fact
a tensor that we can denote $\dpr W^u_O(x)$
verifies:

$$|\dpr W^u_O(x)-\dpr W^u_O(y)|\le D_1 d(x,y)^\b\Eq(3.2)$$

\0 for all $x,y\in W^u_O$ and that there exists $D_2,\b>0$ such
that the jacobian $\dpr S(x)_u$ of $S$ as a map of $W^u_O$ into itself
verify, for all $x,y\in W^u_0$:

$$|\dpr S(x)_u \,(\dpr S(y)_u)^{-1}-1|\le D_2 d(x,y)^\b\Eq(3.3)$$

The analogous conditions are also imposed on $W^s_O$ and on $S^{-1}$.

\0(iii) the property (4) can be put in a quantitative form as follows:
if $dw^u$ and $dw^s$ are two surface elements tangent to $W^u_O$ and
$W^s_O$, respectively, at a common point $x$ then the Liouville volume
of the parallelogram generated by them is
$\m_0(dw^udw^s)=b(x)dw^udw^s$ and $e^{-B_0}<b(x)<e^{B_0}$ holds for
some constant $B_0$.

\0(iv) The continuity above is a "H\"older continuity"; it is
convenient here being generous on the weakness of this assumption and
{\it not} requiring Lipshitz continuity or higher smoothness. Even
very smooth dynamical systems may have just H\"older regularity
of the foliation of phase space into stable or unstable manifolds.

\0(v) The definition essentially yields what is usually called an Anosov
system. A more general notion (of axiom A-system) could be envisaged.
However in the cases considered here the attractors will always be dense
in the full phase space and the above generality will be sufficient.
All the items in the definition are essential; but in a sense the
axiom (1) is the major one among them.
\*

In this section I suppose the system to be chaotic in the above sense.
The following analysis (essentially due to Sinai, Bowen and Ruelle,
[S2], [Bo], [R2]), will show that {\it there is a well defined
statistics for the Liouville measure} and the attractor will be dense
on $\CC$.  The statistics on it can be determined quite explicitly and
it can be shown to have strong ergodic properties.  The existence of a
well defined statistics plays the role of the ergodic hypothesis and
it will therefore be called {\it ergodicity property}: it will appear
that it has been introduced here under the disguise of the chaoticity
assumptions (essentially the smoothness and density of the stable and
unstable manifolds of $O$).

We first describe how the measures $\lis\m^\pm$, the past and future
statistics of the Liouville measure, can be characterized.  An
apparently involved geometric construction is necessary: it is very
simple in the case in which $W^s_O$ and $W^u_O$ are {\it
$1$--dimensional}, hence the phase space $\CC$ has dimension $d=2$. I
shall, however, provide a sketch of the construction in the general
case: but one should first understand it (via drawings) in the trivial
$d=2$ case, to realize later that the general case is its natural
generalization.  The geometrical construction leads to the definition
9 below of {\it Markov pavement}: the construction that follows is
somewhat different from the classical constructions of [S2], [Bo1]
and, admittedly sketchy (except in the $d=2$ case, in which it becomes
a well known construction whose generalization is attempted here): its
details will not be really used in the following and one could invoke
here the theorem of Sinai stating the existence of a Markov pavement,
\ie proceed to the paragraph preceding definition 9 below.

\*
Given a small ball of radius $\d$ (small compared to the curvature
radii of the  manifolds $W^u_O,W^s_O$) centered at $x\in W^u_O$ the
manifold $W^u_O$ will intersect the ball on a (dense) family of
connected surfaces: only one of them will contain $x$ and it will be
called ``the disk on $W^u$ with center $x$ and radius $\d$''. The disks
with radius $\d$ on $W^s$ are likewise defined.

The assumed density and continuity of $W^s_O$ and $W^u_O$ allows us to
define ``disks on $W^u$ or $W^s$'' centered at any point $x$: they are
the surfaces obtained as limits of a sequence of disks of radius $\d$
on $W^s_O$ (or $W^u_O$) centered at the points $x_n\in W^s_O$ (or
$x_n\in W^u_O$) of a sequence of converging to $x$ as $n\to \io$.  And
the union of the disks that match smoothly with a disk on $W^s$ (or
$W^u$) centered at $x$ will define the manifold $W_x^s$ (or $W^u_x$).

Let $\D,\D'$ be two small disks on $W_O^u$ and, respectively, $W^s_O$
centered at $O$.  Let $\th_0$ be so large that the web generated by
$S^{\th_0}\D$ and $S^{-\th_0}\D'$, fills the phase space $\CC$ so densely
that there is no point further away from the web than a prefixed $\d>0$,
and so that the disk of radius $\d$ on $W^s$ centered at any $x\in
S^{\th_0}\D$ has at least another intersection with $S^{\th_0}\D$ (and
``viceversa'', exchanging $\D,\D'$ and $S,S^{-1}$).

On $S^{\th_0}\D$ we consider all the points $x_1,x_2,\ldots$ common to
$S^{-\th_0}\D'$: they form a rather dense set of points on
$S^{\th_0}\D$, and by taking $\th_0$ large we can suppose that there is
a triangulation $\TT$ of $S^{\th_0}\D$ with base on such
points\footnote{${}^6$}{\nota A smooth triangle (one should call it
pyramid but I use the more reassuring word triangle to make the object
look simpler) $T$ is $d$--dimensional smooth manifold $W$ consisting in
a compact set with connected interior containing $d+1$ points, the {\it
vertices}, and a boundary formed by joining them with $d$ smooth
surfaces of dimension $d-1$ each, {\it faces}, containing a different
$d$--ple of the $d+1$ vertices and with the further property that the
set is {\it conical} (see below) around each vertex and with the
intersections of any pair of them forming a smooth $d-1$--dimensional
triangle, {\it face}.  The definition is recursive if one declares that
a $0$ dimensional triangle is a point and a $1$ dimensional triangle is
a smooth arc.  Here conical means that at every vertex one can draw an
open cone: this also requires a definition, as the surface may be not
flat; for instance one can consider a family of smooth curves emerging
from the vertex and tangent at the vertex to a true open right cone with
apex at the vertex and opening angle $\e>0$, and with all the points
close enough to the vertex entirely contained inside the interior of the
triangle.  If the boundaries of the various triangles are required to be
only H\"older continuous rather than smooth one obtains the more general
notion of triangle: the key transversality property remains, however,
unchanged for such more general triangles and we call $\e(T),\h(T)$ the
minimum of the opening angles at the vertices of the various faces of
the triangle and, respectively, the minimum H\"older continuity
exponents of the various faces.  Note that all the $d$ dimensional
triangles have boundaries with $0$ $d$--dimensional volume. A
triangulation of a compact manifold with base on a given family of
points is a covering by a family of triangles with the vertices on the
base, no interior points in common and no base point in their interior.
All this is quite trivial in the case the triangle is $1$ dimensional.
Consider a family of points, distributed on a compact manifold with
smooth boundary, dense enough so that every point has another one closer
than $\d$ to it. Then it is possible to build a triangulation based on
the given points by trimming the manifold boundary by an amount not
exceeding, say, twice $\d$ if $\d$ is small enough. I take this as
obvious, although a proof is likely to be somewhat verbose.} consisting
of smooth triangles $T_j^u$ which have vertices at distances $\le\d$.
By slightly deforming $\D$ we can also suppose that the boundary
$\dpr(S^{\th_0}\D)$ consists of sides of some of the triangles of
$\TT$.\footnote{${}^{7}$}{\nota The deformation is at most of the order
of $\d e^{-\l \th_0}$.} Let $a$ be the {\it smallest distance between
distinct triangles vertices} and let $\e$ be the smallest opening angle
at the vertices.

Let $\dpr_\d^s$ be the surface formed by drawing through {\it each}
$x$ on the {\it boundary} of {\it each} triangle a disk on $W^s$
centered at $x$ and of radius $\d$. This is a codimension $1$ surface.
It is not difficult to see that if the expansion rate $\l$ of $S$ is
large enough (so that $\d D_0 e^{-\l}\ll a$) then we can slightly
modify the sides of the triangles (hence $\D$ as well) so that the
image of $\dpr_\d^s$ is contained in itself.  The expansion rate will,
however, be in general much weaker than needed.  In that case we
imagine, for the time being, replacing $S$ with a high iterate of $S$
so that the new transformation meets the expansivity requirement.

The construction can be done by successive approximations, recursively,
as every change of a given triangle boundary performed to impose the
wanted condition, will ruin the validity of the condition for some other
triangle (but by a smaller amount). One will only be able to infer that
the final triangles have H\"older continuous boundaries, even when the
$W^u_O$ and $W^s_O$ are (locally) flat (it is not difficult to see that
the exponent of H\"older continuity should be bounded below by
$\h=e^{-(\l+-\l_-)}$ if $\l_+$ is the maximal expansion rate and
$\l_-\ge\l$ is the minimal expansion rate on $W^u$, which will be in
general different: just think at what happens to an angle when the
abscissa and the ordinate are contracted by a different scaling factor).
In the case of one dimensional $\D$, however, the construction is
clearly completed ``in one stroke'' and the triangles are just smooth
segments on $W^u_O$ (and their boundaries are isolated points, hence
necessarily smooth ($0$-dimensional manifolds).  The reason why in
general the construction leads to triangles with mildly regular
boundaries can be understood through the celebrated elementary Bowen's
example, [Bo2], ultimately relating it to the direction dependence of
the expansion rate on the unstable manifold, when it has dimension $>1$
and at least two distinct positive Lyapunov exponents (so that
$\l_+>\l_-$ as expected from the above bound on the H\"older continuity
exponent).  The deformation of the triangles, during the construction,
is small (at most of the order of $\d e^{-\l \th_0}$) but the regularity
of the deformation is quite out of control (because of the mechanism in
[Bo2]).

An identical construction can be performed on the stable manifold
leading to a triangulation $\TT^s$ of $S^{-\th_0}\D'$, with H\"older
cotinuous triangles.

If $\TT^u$ and $\TT^s$ are the above triangulations of $W^u_O$ and
$W^s_O$ we can consider the pairs of triangles $T^u_h,T^s_k$ with one
point in common and consider the sets $E_j=T^u_h\times T^s_k$ built
with the points obtained by fixing $\x\in T^u_h$ and $\h\in T^s_k$ and
by drawing the disk with radii $2\d$ on $W^u_\h$ centered at $\h$ and
on $W^u_\x$ centered at $\x$ and by considering their intersection
$x=x(\x,\h)$ as $\x,\h$ vary in $T^u_h$ and $T^s_k$, (here $\d$ may
have to be required to be small enough, always with respect to the
curvature radii of the manifolds, so that the intersection defining
$x$ is certainly non empty).

Such sets will be called {\it prisms}: the name "parallelogram" or
"rectangle" is used in the literature: but the construction here is
somewhat different from the original one of [S2],[Bo] (which is not
based on triangulations).  Furthermore even in the original construction
the sets called parallelograms look to me, no matter how I attempt to
draw them, prisms except of course in the two dimensional case, when
they really look like parallelograms. The sets
$T^u_h$ and $T^s_k$ will be called, respectively, the {\it horizontal}
and the {\it vertical axis} of the prism $E_j=T^u_h\times T^s_k$.
Therefore a natural name for the sets $\dpr^u E_j\= T^u_h\times \dpr
T^s_k$ and $\dpr^s E_j\=\dpr T^u_h\times T^s_k$ will be {\it horizontal}
boundary of the prism $E_j$ and, respectively, {\it vertical boundary}
of $E_j$ (or unstable and stable boundary of $E_j$).  The set
$\dpr^s=\cup_j
\dpr^s E_j$ $\dpr^u=\cup_j \dpr^u E_j$ define the {\it total vertical}
or the {\it total horizontal boundary} of the family of prisms $E_j$
which themseves constitute a {\it pavement} of the phase space with
small {\it prisms}.

The above notion of prism as a set having the form $E=T^u\times T^s$,
see the above two last papragraphs, where $T^u$ and $T^s$ are small
connected surfaces, with H\"older continuous boundaries, on the stable
and unstable manifolds of a common point $x$ is more general; so that
one can consider a pavement $\EE$ of the phase space $\CC$ with prisms,
$\EE=(E_1,\ldots)$ and define, likewise, the stable and unstable
boundaries $\dpr^u_j,\dpr_j^s$ of the prisms $E_j$ and the total
horizontal and the total vertical boundaries $\dpr^u$ and $\dpr^s$: note
that $\dpr^s,\dpr^u$ have zero volume (by the H\"older continuity of the
boundaries). Given such concepts the following definition is useful:
\*

{\bf 9 Definition:} {\it A pavement $\EE=(E_1,E_2,\ldots)$ by prisms is
called a {\sl Markov pavement} (or a {\sl Markov partition}) if the total
stable boundary is mapped into itself under the action of $S$ and the
total unstable boundary is mapped into itself under the action of $S^\1$
and they have both $0$ volume.} \*

The key property of the special pavement $\EE$ constructed above, and
motivating the last definition, is that by construction the total
vertical boundary of the prisms is mapped into (a small portion of)
itself under the action of $S$ and likewise the total horizontal
boundary $\dpr^u$ is mapped into (a small portion of) itself under the
action of $S^{-1}$. Hence we have just seen that special Markov
partitions, associated with the stable and unstable manifolds of a
hyperbolic fixed point, exist for a chaotic system at least if $S$ is
replaced by a high enough power $S^K$ of $S$. Such special Markov
partitions are particularly suited for numerical applications, see [FZ].

But it follows immediately from the definition 9 just given (and the
covariance of $W^s_x,W^u_x$: $S W^s_x=W^s_{Sx}$, $S W^u_x=W^u_{Sx}$)
that if $\EE$ is a Markov partition for $S^K$ then the partition
$S^p\EE$ (whose prisms are obtained by transforming with the map $S^q$
those of $\EE$) is {\it also} a Markov partition for the iterate $S^K$
of $S$. Furthermore the partition obtained by intersecting the
prisms of the partitions $S^p\EE$ with $p$ taking $K$ consecutive
values is a Markov partition for $S$. Hence any dynamical system which
is chaotic in the sense of definition 8 admits a Markov partition
$\EE$, [S2].

We can form, given any integer $\th$, and a Markov partition $\EE$ a
more general and finer Markov partition $\EE_\th$ simply by
"intersecting" the partitions $S^{-\th}\EE,S^{-\th+1} \EE,\ldots,
S^{\th} \EE$.  We shall only consider sequences of Markov partitions of
the above form $\EE_\th$.

Given $\EE$ we consider the map $\Si:\ x\to\V\s(x)$ mapping $x$ to the
sequence $\s_j$ defined by $S^jx\in E_{\s_j}$, the {\it history} of $x$
on the partition $\EE$.

{\it This map is unambiguously defined for all $x$ such that $S^kx$ does
not fall on the boundary of any one of the parellelograms $E_\s$ of
$\EE$}.  If we define the {\it compatibility matrix} element $M_{\s\s'}$
to be $1$ if $SE_\s^0\cap E_{\s'}^0\ne\emptyset$, and $0$ otherwise,
where $E^0_\s$ is the set of interior points of $E_\s$, then it is clear
that the history $\V\s(x)$ of a point $x$, for which it is unambiguously
defined, is an {\it allowed sequence}: in the sense that
$\prod_{j=-\io}^\io M_{\s_i,\s_{i+1}}=1$.  Viceversa, and this follows
directly from the mappimng properties of the boundaries of the prisms in
definition 9 (attempt a proof only after finding via a drawing in the
trivial $d=2$ case), if $\V\s$ is an allowed sequence there is always a
point $x$ producing $\V\s$ as history, provided one chooses conveniently
the definition of $\s_j(x)$ in the cases in which this quantity is
ambigously defined (because some iterate of $x$ falls on the boundary of
some $E_\s$).  Item (5) of the chaoticity assumption, in definition 8,
implies that the matrix $M$ is ``transitive'', \ie there is an integer
$p$ such that $M^p$ has all entries non vanishing, [S2]. This important
property may be implied by other more topological properties, see the
discussion of the axiom--A in [Sm], but here we do not discuss this
point (in the same spirit animating what precedes and follows, of not
trying to get ``optimal theorems'' operating with ``minimal
assumptions'').

If $\V\s$ is the history of $x$: $\V\s=\Si x$, then the history of
$Sx$ is $\f\V\s$ where $\f$ is the {\it shift} to the left of $\V\s$.
This means: $\Si(Sx)_j=(\Si x)_{j+1}$.

Even when the history of a point is ambiguous it can be seen that the
number of possible {\it allowed} sequences is never more than a number
that can be explicitly bounded in terms of the maximum number of
triangles with one vertex in common and the dimensions of $W^s,W^u$.
But we do not have to worry about this, since the boundaries of the
prisms of $\EE_\th$ have zero volume measure and we are interested
only on properties valid for the data $x$ which have probability $1$
with respect to the Liouville measure.

Therefore, for our purpose, we can think that the points of $\CC$ can be
described by allowed (in the above sense) sequences of symbols $\V\s$:
{\it we shall denote $K$ the space of such sequences}, defining on it a
metric that sets to $2^{-q}$ the distance between two sequences which
agree on the sites $-q,\ldots,q$ but not on any larger symmetric
interval.  Here $2$ is arbitrarily chosen and any number $>1$ could
replace it, generating the same topology on the space of sequences.  In
other words we can use the sequences $\V\s$ as a system of coordinates
(a slight variation on the usual representation of the cartesian
coordinates in some digital representation of the reals, \eg base $10$;
but the present representation is intrinsically tied to the dynamics).

It is convenient to think of a history sequence
$\V\s=(\s_j)_{j=-\io}^\io$ as a {\it configuration of a one dimensional
spin system}, to make more striking the analogy (that will emerge below)
with the theory of one dimensional statistical mechanics : thus we call
the labels $j$, labeling the times marking the history (in units of
$t_0$), with the name of {\it sites}.  Hence the value of $\V\s$ at the
site $j$ will be $\s_j$.

The Liouville measure $\m_0$ becomes a probability distribution
$\m'_0$ on the space of the allowed sequences. In fact consider the
set of sequences denoted

$$G'=\pmatrix{-q&.&.&.&q\cr\s^0_{-q}&.&.&.&\s^0_q\cr},$$

\0consisting in the sequences $\V\s$ whose values $\s_j$ with
$j=-q,\ldots,q$ coincide with the given $\s_j^0$. Such set $G'$ will be
naturally given a probability equal to the $\m_0$--measure of the set:
$G=\Si^{-1}G'$, if $\Si$ is the above history map.

Thus setting $\m'_0(G')\=\m_0(G)$, with $G=\Si^{-1}G'$, for any
(Borel) set $G'$ in the space $K$ allows us to
think of the dynamical system $(\CC,S)$ as $(K,\f)$ and the probability
measure $\m_0$ becomes $\m_0'$.

{\it The main point is that there is a simple formula expressing the
probability $\m'_0(G')$}, (Sinai).

To find it just remark that the set $G$ is (clearly) a prism for the
Markov partition $\EE_{q}$ (because $G\=\cap_{r=-q}^q S^{-r}
E_{\s^0_r}$).  If $\b^s_\s$ denotes the surface of the vertical axis
$T^s_\s$ of $E_\s=T^u_\s\times T^s_\s$ lying on $W^s_O$ then the area of
the vertical axis of $G$ is essentially given by:
$\b^s_{\s_{-q}}\prod_{j=-q}^{-1}\L_s(S^jx)$, where $x$ is a point in $G$
and $\L_c(x)$ is the absolute value of the jacobian determinant of $S$,
at $x$, as a map from $W^s_x$ to $W^s_{Sx}$.

Likewise if $\b^u_\s$ is the surface of the horizontal axis $T^u_\s$ of
$E_\s$ on $W^u_O$ then the area of the horizontal axis of $G$ is,
essentially: $\b^u_{\s_{q}}\prod_{j=0}^{q-1}\L_u^{-1}(S^jx)$ where
$\L_u(x)$ is the absolute value of the jacobian determinant of $S$, at
$x$, as a map from $W^u_x$ to $W^u_{Sx}$.

Therefore the $\m_0$ measure of $G$ is, essentially:

$$\b^s_{\s_{-q}}\Big(\prod_{j=-q}^{-1}\L_s(S^jx)\Big)\cdot
\Big(\prod_{j=0}^{q-1}\L_u^{-1}(S^jx)\Big)\b^u_{\s_{q}}\cdot
b(x)\Eq(3.4)$$

\0where $b(x)dw dw'$ is the volume element (\ie the
$\m_0(dwdw')$ corresponding to two surface elements $dw,dw'$
tangent to, respectively, $W^u_x,W^s_x$ (see remark (iv) above).
The "essentially" means that there is an error in \equ(3.4) due to the
finite size of the sides of $G$: see below for its treatment.

To connect \equ(3.4) with something more familiar we can go through
the definitions and remarks  that led Ruelle, [R1],[R3], to call what
follows the {\it thermodynamic formalism} for strange attractors.
Define:

$$\eqalign{
\txt h_+(\V\s)=&\log \L_u(x(\V\s)),\kern.6truecm
h_-(\V\s)=-\log \L_s(x(\V\s))\cr
h_s(\V\s)=&\log \b^s(\s_0),\qquad h_u(\V\s)=\log\b_u (\V \s_0),
\qquad h_0(x)=-\log b(x(\V\s))\cr}\Eq(3.5)$$

\0and remark that $h_0,h_+,h_-$ as functions of $\s_j$ depend very little
on the $\s_j$'s with $j$ large in the sense that ther exist constants
$H_\g, \k>0$ such that for $\g=u,s,+,-,0$:
$$|h_\g(\V\s)-h_\g(\V\s')|<H_\g e^{-\k q}\Eq(3.6)$$

\0if $\s_i=\s'_i$ for $|i|<q$.
This follows again from the the continuity and hyperbolicity assumption,
because varying $\V\s$ on the sites $j>k$-th means varying $x(\V\s)$ in
$$\Si^{-1} \pmatrix{-k&.&.&.&k\cr\s_{-k}&.&.&.&\s_k\cr}$$ which is a
prism whose diameter has size of order $O(\d e^{-\l k})$, see
\equ(3.1).  Therefore, by the continuity assumption in definition 8, we
see that the variation of $h_j$, $j=0,\pm,u,s$, is bounded by $O(e^{-\l
k\b}\d^\b)$.  This is usually quoted by saying that dependence of
$h_j$ on $\s_k$ ``vanishes exponentially'' or $h_j$ has ``memory vanishing
exponentially''.

In terms of the functions $h_\g, \g=s,u,+,-,0$ we can rewrite
\equ(3.4) as:

$$e^{-h_s(\f^{-q}\V\s)-\sum_{j=-q}^{-1}
h_-(\f^j\V\s)-h_0(\V\s)-
\sum_{j=0}^{q-1}h_+(\f^j\V\s)\,-h_u(\f^q\V\s)}\Eq(3.7)$$

This shows, unless the approximation involved in the discussion of the
word "essentially" above spoils everything, that the probability
distribution $\m_0'$ on $K$ coincides with a Gibbs' state on $K$ for
the short range {\it non translation invariant} formal hamiltonian:

$$\sum_{j=-\io}^{-1} h_-(\f^j\V\s)+h_0(\V\s)+
\sum_{j=0}^\io h_-(\f^j\V\s)\Eq(3.8)$$

\0Such a state, to the far right (\ie in the far future) corresponds to the
Gibbs state $\lis\m_+'$ with hamiltonian $\sum_{-\io}^\io h_+(\f^j\V\s)$
and to the far left (\ie in the far past) corresponds to the Gibbs state
$\lis\m_-'$ with hamiltonian $\sum_{-\io}^\io
h_-(\f^j\V\s)$.\footnote{${}^8$}{\nota One should not forget that
\equ(3.8) is defined for the allowed sequences. Hence one should more
precisely say that the hamiltonian of our spin system is
\equ(3.8) if the configuration is allowed and it is $+\io$ otherwise:
\ie the spin system has a (short range) {\it hard core} coresponding to
the compatibility condition.}

The last statement defines unambiguously $\lis\m_\pm'$ as probability
measures on $K$, via the theory of one dimensional Gibbs states
([R5],[Bo],[Ga3]). This is so because the Gibbs states potentials have
short range (by the above remark on the shortness of the memory, the
potential decreases exponentially, see the paragraph following
\equ(3.5)), and no phase transitions are possible, the system being
one dimensional. By the map $\Si$ the distributions $\lis\m_\pm'$ are
transformed into the natural candidates for the forward and backward
statistics $\lis\m_\pm$ for the distribution $\m_0$.

Hence the only problem is the check that $\m_0'$ is really the Gibbs
state on the allowed sequences with formal energy \equ(3.8): this
means discussing quantitatively the error mentioned in connection with
\equ(3.4). All the properties of $\lis\m^\pm$ would then follow from
the well known theory of the one dimensional short range Gibbs states
for spin systems, [R5],[Bo1], including the property that the
$\lis\m'_\pm$ are the forward and the backward statistics of $\m'_0$.
The latter property holds simply because the Gibbs state with
hamiltonian \equ(3.8) really converges to $\lis\m'_+$ or to $\lis\m_-$
if it is observed to the far right or to the far left.  Thus the
characterization of $\lis\m^\pm$ would have received a complete
description.

To correct the error involved in the above use of the word "essentially"
one can use the deeper properties of Gibbs states; namely their
characterization in terms of the DLR equations.  To use the DLR theory
one needs an expression of the ratio of the probabilities that two
sequences $\V\s^1$ and $\V\s^2$ have the value $\s$ or $\s'$ at site
$0$, conditioned to the two sequences having the same value in all the
other sites.  This is the ratio of the probabilities for the events in
which the spin at site $0$ is $\s$ or $\s'$, {\it having fixed} the
same configuration for the spins at {\it all} the other sites.

This means considering the ratio of the probabilities:

$$\lim_{q\to\io}
\fra{\m_0'(\pmatrix{-q&.&.&-1&0&1&.&.&q\cr\s_{-q}
&.&.&\s_{-1}&\s&\s_1&.&.&\s_q\cr})}
{\m_0'(\pmatrix{-q&.&.&-1&0&1&.&.&q\cr\s_{-q}
&.&.&\s_{-1}&\s'&\s_1&.&.&\s_q\cr})}\Eq(3.9)$$

\0where $\V\s=(\s_j)_{-\io}^\io$ is an arbitrary (allowed) sequence.
Then clearly the error involved in \equ(3.4) can be regarded as a
multiplicative factor correction to the ratio in \equ(3.9) approaching
$1$ as $q\to\io$.  If $\V\s^1$ and $\V\s^2$ denote two allowed sequences
differing only in the entry with label $0$, one gets that the value of
the limit in \equ(3.9) is, rigorously:

$$e^{-\sum_{-\io}^{-1}(h_-(\f^j\V\s^1)-h_-(\f^j\V\s^2))
-(h_0(\V\s^1)-h_0(\V\s^2))-\sum_{0}^{\io}(h_+(\f^j\V\s^1)-h_+(\f^j\V\s^2))
}\Eq(3.10)$$

\0which is the DLR characterization of $\m'_0$ as the Gibbs state
with formal hamiltonian \equ(3.8): proving, by the uniqueness theory of
the short range one dimensional Gibbs states and the DLR theory of the
Gibbs states, that indeed $\m'_0$ {\it is the Gibbs state with energy
\equ(3.8)}, [Bo1],[Ga3].  \*

{\it Hence the forward and backward statistics $\lis\m^\pm$ are
precisely identified and described}, because the theory of one
dimensional short range Gibbs states is very well developed.  {\it It
also follows that $\m_\pm$ are ergodic (in fact isomorphic to a
Bernoulli shift) measures}, (see [Ga2]).
\*

We proceed to derive some properties of $\lis\m^+$ by making full use of
the short range \equ(3.6) of the potential $h_+$ generating $\lis\m^+$
as a Gibbs state: they will be used in the application in \S4.

Let $\G_T, T>0$ denote the set of the allowed strings $\s^T_j, |j|<T$:
these are the  strings that can becontinued to infinite allowed strings
$\hat{\V\s}^T\in K$: we in fact imagine to continue each ${\V\s}^T$
to some $\hat{\V\s}^T$ (arbitrarily).

Then $\lis\m^+$ can be constructed as a limit of the distributions
$\m_T$ defined by the average they assign to an arbitrary smooth
function $F$:

$$\ig \m_T(dx) F(x)=\fra {\sum_{{\V\s}^T\in \G_T}e^{-\sum_{-T}^T
h_+(\f^j\hat{\V\s}^T)} F(x(\hat{\V\s}))}{Z_{[-T,T]}}\Eq(3.11)$$

\0where $Z_{[-T,T]}$ is the normalization factor,that should be called
the ``partition function'' of the energy $h_+$ relative to the interval
$[-T,T]$.

Since $\lis\m^+$ is the Gibbs state with potential $h_+$ it follows that
$\lis\m^+$ is the limit of $\m_T$ as $T\to\io$, [S2]. We now investigate
the probability distribution of a random variable $w=\sum_{j=-t}^t
f(S^jx)$ with $f$ smooth and $t>0$. This is done by considering the
probability distribution $\lis\m_{T,t}$ defined, on the smooth functions
$F$, by:

$$\ig \m_{T,t}(dx) F(x)=\fra{\sum_{{\V\s}^T\in \G_T}e^{-\sum_{-t}^t
h_+(\f^j\hat{\V\s}^T)} F(x(\hat{\V\s}))}{\rm normalization}\Eq(3.12)$$

\0We want to show:
\*
\0{\bf Theorem II:} {\it If $f$ is smooth the probability distribution
of $F=\sum_{-t}^t f(S^jx)=w$ computed by using the distribution
$\lis\m_{T,t},\,T>t$ is different from that obtained by using $\lis\m^+$
by a factor bounded, for each value of $w$,  between $e^{-B_f}$ and
$e^{B_f}$  for some $B_f$ which is $T,t$ independent.}
\*

\0{\it Remark:} this is a simple consequence of the short range
(\equ(3.6) of $h_+$ and the genral theory of Gibbs states on one
dimensional lattices. Note that the set of strings in $K$ (the space of
the compatible strings) which agree with a given $\V\s^T\in \G_T$
is a prism of the Markov partition $\EE_T=\cap_{k=-T}^T S^k\EE$; so that
the sum in \equ(3.12) can be regarded as a sum on the prisms of $\EE_T$
\*
\0{\it partial proof}: since $f$ is smooth the function
$f'(\V\s)=f(x(\V\s)$ also has exponentially vanishimng memory, as the
$h_\g$ in \equ(3.6): hence there exist $C,\n$ such that
$f'(\V\s)-f'(\V\s')|<C e^{-\n q}$ if $\V\s,\V\s'$ agree on the sites
between $-q$ and $q$. Therefore we suppose first that $f'$ depends {\it
only} on the sites $j\in[-r,r]$, (``finite range assumption''). The the
value of $w=\sum_{-t}^t f(S^jx(\V\s))$ is entirely determined by
$\V\s^T_j, \,|j|<t+r$, so that the probability of $w$ can be computed
with the distribution:
$$e^{-\sum_{j=-t}^t h_+(\f^j\V\s^T)}.
\fra{\sum_{{\V\s}^T\in \G_T\atop\V\s^{t+r}\,{\rm fixed}}
e^{-\sum_{j=-T}^{-t-1} h_+(\f^j\V\s^T)}
e^{-\sum_{j=t+1}^{T} h_+(\f^j\V\s^T)}}{Z_{[-T,T]}}\Eq(3.13)$$

\0but since $h_+$ has short range the above numerator is essentially the
product of the partition functions relative to the intervals $[-T,-t-1]$
and $[t+1,T]$, with some $\V\s^t$ dependent ``boundary condition'' (in
the sense of the Gibbs states terminology). Therefore the ratio is
bounded {\it between}:

$$e^{-\sum_{j=-t}^t h_+(\f^j\V\s^T)}.
\fra{Z_{[-T,-t-1]}Z_{[t+1,T]}}{Z_{[-T,T]}}. e^{\pm B_r}\Eq(3.14)$$

\0where $B_r$ is a $T,f$--independent constant: note that the ratio of
the partition functions is large or small (exponentially in $T,T$) but
{\it it is not random} (\ie it is independent of $\V\s^{t+r}$).

This proves the theorem if $f$ has finite memory $r$: if $f$ has
exponentially vanishing memory the result has to be proven by
approximating it with a finite memory function and passing to the limit;
I do not discuss this problem further.
\*

\*
\0{\it\S4 Time reversible systems. Ruelle's principle.}
\numsec=4\numfor=1
\*
As an application I consider the system in example 3, \S1, to show that
the analysis of \S3 can have some far reaching consequences and
relevance for a fundamental theory of non equilibrium ensembles.
Consider a $2$ dimensional periodic box containing a few circular
obstacles disposed so that there are no collisionless trajectories (this
is the case called $0H$ in [GG], "no horizon").  The box contains $N$
particles and the equations of motion are the \equ(1.1), \ie with the
usual symbols:
\ifnum\mgnf=1
$$\eqalign{
\dot\qq_j=&\fra1m \V p_j\cr
\dot\pp_j=&-\Dpr_{\qq_j} V_{ext}(\qq_j)-\Dpr_{\qq_j}
V(\qq_1,\ldots,\qq_N)+
E \V i-\a(\pp)\pp_j\cr}\Eq(4.1)$$
\else
$$
\dot\qq_j=\fra1m \V p_j,\quad
\dot\pp_j=-\Dpr_{\qq_j} V_{ext}(\qq_j)-\Dpr_{\qq_j}
V(\qq_1,\ldots,\qq_N)+
E \V i-\a(\pp)\pp_j\Eq(4.1)$$
\fi
where $V_{ext}$ is the hard core potential (\ie it is really a boundary
condition imposing elastic reflections on the obstacles) and $V$ is a
short range pair potential energy. The $\a$ represents a friction
mechanism to keep the total energy $H=\sum_i\pp_i^2/2m+
V(\qq_1,\ldots,\qq_N)$ bounded. We choose: $\a(\pp)={E\V
i\cdot\sum_j\pp_j}/{\sum_j\pp_j^2}$ which keeps $H$ constant, exactly.
This is called a "gaussian thermostat" from the meaning it has when
$V=0$.  The case $V=0, N=1$ has been "completely" studied in
[S1],[CELS].

The dynamical system generated by \equ(4.1) can be regarded as defining
a dynamics on the $4N$ dimensional full phase space $\FF$, or on the
$4N-1$ dimensional surface of constant energy $\EE$, or on the $4N-2$
dimensional manifold $\CC$ consisting in the phase space points of $\EE$
in which one particle is exactly colliding with one of the hard
obstacles.  We denote $S^0_t$ the dynamics on $\FF$, $S_t$ that on $\EE$
and $S$ will denote the map defined, on $\CC$, by mapping one collision
to the next.

The system properties are:
\*
\0{\bf (A) Dissipativity}: the phase spaces $\FF,\EE,\CC$ contract at an
average rate $(2N-1)\media\a$, where $\media\a$ is the time average of
$\a$, at least if we suppose that the forward infinite time average
$\media\a$ is positive, as it seems intuitively correct and as some
numerical experiments seem to support.
\*

\0{\bf (B) Reversibility}: the map $i: (\qq,\pp)\to(\qq,-\pp)$ is such
that, if $t\to x(t)$ is a solution of \equ(4.1), then $t\to
ix(-t)\=(ix)(t)$ is also a solution.
\*
I will furthermore suppose that the system verifies the following
property:

\*
\0{\bf (C) Chaoticity}: in the sense that ``things go as if the system
was chaotic".
\*

The strict validity of this assumption in the sense of definition 8,
\S3, is {\it not true} for the model \equ(4.1) already for $N=1$,
simply because the system is {\it not} smooth.  For large $N$ testing
its {\it approximate} validity is very hard, but hopefully something will be
done in the future.  The best one can hope is that the failure to be
verified is not "relevant" for the discussion of the thermodynamics of
the system, in the same sense that the failure of ergodicity is believed
to be, in many cases, irrelevant for the equilibrium thermodynamics
(\ie in the limit as $L\to\io$ with fixed particle density and
periodically repeated obstacles).

This is essentially Ruelle's principle: its operational meaning is to
proceed as if the system was chaotic in the sense of the definition 8 of
\S3, or some similar definition, and then to suppose that the
deductions are correct even though we have no way to check the
chaoticity assumption or even though it does, strictly speaking, fail (as
in the case at hand, \equ(1.1)).  One can hope that the assumptions
should become more and more "true" as $N$ grows.  They fail, as just
mentioned, for instance if $N=1$, see [CELS]: but, in this particular
case, some conclusions of \S3 would nevertheless be correct {\it
even if $N=1$}, as it seems likely to be implied by combining the
present analysis with that of [CELS].

For instance an example of a property that is expected to hold, if the
chaoticity assumption holds, is a strong, exponential, decay of the
correlations between smooth observables, [S2], [R2], [Bo1].  The above
assumption is therefore expected to correspond to a strong exponential
decay of the correlations, at least over a time scale that is reachable
by the experiments that can be conceived in order to check the picture
that we are developing.

The latter decay has not been experimentally investigated (because it is
really difficult).  But after a decade of uncertainty it is becoming
increasingly acceptable that indeed the correlations, in the $N=1$,
$E=0$ case, may decay exponentially: in the paper [GG] the "simple"
cases $V=0,N=1$ and several choices for $V_u$ are studied.  No evidence
for a non exponential decay could be detected, at least on time scales
reliably attainable by numerical experiments (and, of course, for the
simple observables considered).  Essentially all the data confirm,
instead, an exponential decay, both for the observables of the collision
system $(\CC,S)$ and for the continuous system $(\EE,S_t)$.  In this
respect "things go as if the system was chaotic".\footnote{${}^{11}$}{\nota
In our paper there is only one experimental result that can rise, in our
opinion, doubts about the decay law.  It is the result corresponding to
the extremely long run leading to fig.  6 of [GG]: see the discussion
there. More recent work, [ACC], provides much stronger further evidence
on the exponential decay.} In spite of the evidence for the failure of
the assumption, we have therefore an example, in the case least
favourable for thermodynamic interpretations (as $N=1$), strengthening
the view that "things may go as if the system was hyperbolic" expressed
here.

The existence of time reversal invariant periodic orbits can be
deduced for many arrangements of the obstacles: hence we shall also
suppose, for simplicity that the periodic point discussed in the
definition of chaoticity is just a time reversal invariant periodic
point. And to simplify further the analysis we assume that in fact $O$
is a fixed point (it should be clear that the analysis that follows
would not really change if $O$ was just a periodic motion).

Other properties can be expected to hold for our model, on the basis
of analogies with similar models in which they have been shown to
hold.  I make here a list, hoping that someone will find their
experimental check worth of some effort. They {\it will not bee used
in the following} but they make the general picture brighter.

\* {\it (D) Density}: following [LPR] we label the non negative
Lyapunov exponents of $S_t$ (on $\EE$) as $\l^+_j$,
$j=2N-1,2N-2,\ldots,n_+$, in decreasing order, and the negative ones as
$\l^-_j$, $j=2N-1,\ldots,n_-$ in increasing order, discarding from the
enumeration the trivially zero exponent associated with the flow
direction.  Thus the maximum Lyapunov exponent is
$\l_{\max}\=\l^+_{2N-1}$.  Then in various models it has been
experimentally\footnote{${}^{12}$}{\nota By {\it experimental
measurement} I mean here also a computer experiment: as I think it does
not conceptually differ from the older notions of experimental
measurement.} found, starting with the work [LPR] which has been
followed by other interesting cases,(\eg [ECM1], [SEM]), that the graphs
of $x=j/(2N-1)\to \l^\pm_j$ seem to have a smooth limit $f_\pm(x)$ as
$N\to\io$ and $x>x^\pm$.  No experimental result seems to be available
in the case \equ(4.1): I shall assume this property to hold for it.  The
Lyapunov exponents of $S$ (on $\CC$) will be $t_0\l^\pm_j$, where $t_0$
is the average time between any two collisions, which is the time scale
associated with the map $S$).  \*

{\it (E) Pairing}: it has been noted that in various cases, starting
with [Dr],\-[ECM1], that:

$$\l^+_j+\l^-_j=j-{\rm independent}= -\media\a\qquad
j=2N-1,\ldots\Eq(4.2)$$

\0Where $\media.$ denotes (forward) time average (over $\io$ time).

The relation \equ(4.2), that will be called {\it strong pairing
property}, has been experimentally found to hold for many other models,
as it emerges from the subsequent analysis in [SEM]. It is not yet clear
whether it holds strictly or up to corrections of $O(N^{-1})$.

\* \0{\it Remarks:} \* \0(1) The pairing property appears to be
related to several features of the equations of motion among which the
reversibility.  As essentially suggested in [Dr] one may think that
the property holds, for instance, for locally hamiltonian equations
(as \equ(4.1) with $\a=0$) with a constraint imposed via a variational
principle (like the Gauss' least constraint principle or some
generalizations of it).  But no proofs are available and, furthermore,
the case [ECM1] and others in [SEM] do not fall into this cathegory,
although \equ(4.2) holds for them.

\*
\0(2) In other models considered in [SEM] a weaker property seems to hold:

$$-C\media\a\le\l^+_j+\l^-_j\le0\Eq(4.3)$$

\0for some $C>0$, that I will call the {\it weak pairing
property}, whose essential feature is the $N$--independence of
$C$.\footnote{${}^{13}$}{\nota In [SEM] the Lyapunov exponents are
considered in the full phase space $\FF$.  As discussed in [SEM] the
rules \equ(4.2) or \equ(4.3) may fail to be obeyed in a few cases: but
this seems often related to the way the exponents are counted and
computed.  For instance in some systems there may exist observables
called in [SEM] "constants of motion" or "conservation laws" which, if
initially having some special value, conserve it in the evolution, (note
that this is a somewhat unusual notion of conservation law, as it holds
only for special initial values), or have slow oscillations around it .
For instance $H$ or the flow direction in \equ(4.1), or the total
momentum $\V P$ in the case $V_{ext}=0$ (in the latter case this is
conserved only if $\V P=\V0$).  And one is naturally led to fix the
values of such observables or to perform the measurements in some
special way (\eg by timing them appropriately when there is a periodic
forcing).  This is done to simplify the calculations, eliminating
trivially behaving coordinates: but the end result might be in a odd
number of exponents and in a consequent apparent "failure" of the
pairing rule.} Neither \equ(4.2) nor \equ(4.3) have been tested
numerically for the present model.  I consider quite likely that the
strong pairing rule holds in this case.
\*

\0(3) The pairing property and the density property imply that the
Kaplan Yorke fractal dimension of the attractor (also called the
Lyapunov dimension), see [ER], is {\it macroscopically smaller} than
the full dimension $4N-4$ of the phase space, by an amount
$O(\l_{\max}^{-1}\media\a N)$: the proportionality to $N$ of the loss
in dimension is sometimes called ``dimensional reduction'', [PH],
[CELS].

\* Going back to the system \equ(4.1), after the assumption (C) above,
I want to discuss its consequences, with the main objective of finding
some that could be tested and thus provide a test for Ruelle's
principle for the statistics $\lis\m$, and to see if any experimental
consequences can be drawn from it (accessible by using the present day
technical capabilities).

Under the assumptions (C) the "principle" is in fact a {\it theorem}
(see theorem II,\S3): the reason it is called a principle is that, as
already mentioned, the hypotheses of definition 8 are not to be
expected to hold rigorously for our model: they must be regarded to
hold approximately, and this is the reason we say that to use them is
a "principle", which we regard as correct for all the consequences it
may have about thermodynamic quantities.

Note first that the time reversal symmetry implies that $W^u_O$ and
$W_O^s$ have the same dimension, equal to half that of the phase space,
\ie $2N-1$.

A key idea behind all what is said in this paper, and on which intuition
can be built, [Ga1], is that of regarding $A$, rather than as a fractal
set of $\sim4N$ dimensions, as a {\it smooth surface} of dimension
$2N-1$ consisting in the unstable manifold $W_O^u$ of some periodic
point $O$.  Note that, in this way, there is unification between the
equilibrium ($E=0$) and the non equilibrium cases (in both cases $W_O^u$
is smooth and it has the same dimension).

One should avoid thinking of $A$ as a nasty fractal: following [Ga1] one
must think of it as an infinite uncut {\it folio} (\ie $W_O^u$) confined
into a finite region (the phase space $\CC$) and folded over and over
again to fit into it, thus forming the (uncut) "book" $A$. Forcing simply
introduces some (small if $E$ is small) wrinkles on the folio
accounting for the (mild) fractal nature of $A$.  In particular the
dimension of $W_O^u$ stays $2N-1$ and does not change because of the
introduction of the forcing (while the dimension of $A$
changes by $O(N)$).

The unification between the equilibrium theory and the non equilibrium
one is made possible by regarding the attractor in the phase space $\CC$
as $2N-1$ dimensional (rather than $\sim 4N-2$ dimensional as one would
be normally tempted to do).

{\it The real question is whether, if $E\ne0$, we can predict anything}
that could be experimentally checked as a test of Ruelle's principle.

The work [ECM2] provides evidence in this direction. I apply here the
ideas of [ECM2], [CG], to show their relevance to the present case (as
an example of the claim made in [CG] about the generality of the ideas).

We can study the fluctuations, in the stationary state
$\lis\m$, of the average $(2N-1)\media{\a}_t$ of the phase space
contraction rate $(2N-1)\a$ over time stretches $t\, \t_0$:

$$\media{\a}_t(x)=\fra1{t \t_0}\ig_{-t\, \t_0/2}^{t \t_0/2}\a(S_\th x)d\th=
\fra1{t\, \t_0}\sum_{j=-t/2}^{t/2-1}
\ig_0^{\t_0(S^jx)}\a(S_\th S^jx)\,d\th\Eq(4.4)$$

\0where $ \t_0$ is the average time elapsing between timing events, \ie the
average (over infinite time) of the time interval $\t_0(x)$ between the
event $x$ and the successive $Sx$; the $S_\th$ is the continuoum time
evolution on the surface of constant energy.

The quantity $(2N-1)\media{\a}_t$ is also called the {\it entropy
production rate} on the trajectory stretch between $S^{-t/2}x$ and
$S^{t/2}x$: it is related to a transport coefficient (the conductivity,
\ie to the ratio between the particle current and the field $E$, see
[CELS]), at least in the present model (because the relation between
entropy production and a suitable transport coefficient might only hold
for a special class of examples, see [CL]).

If the quantity $t$ is small compared to the duration of the experiment
the quantity $\media{\a}_t$ {\it fluctuates} around the mean value
$\media\a$ (defined as the (forward) average of $\a$ over infinite time).
If we write:

$$\media{\a}_t(x)=\media\a\, a_t(x)\Eq(4.5)$$

\0we define a dimensionless
random variable $a_t(x)$ with (forward) average $1$.  We can divide the
axis $a_t$ into intervals $I_{j\d}$, $j=0,\pm1,\ldots$, with $I_{j\d}=
[j\d,(j+1)\d]$ for some small $\d$ and measure the probability
distribution of $a_t$ by counting how many times $a_t$ takes a value in
each $I_{j\d}$ when measured at the phase space points into which $x$
evolves at times multiples of $t\,\t_0$,

The measurement of $\media{\a}_t(x)$ requires measuring the value of
$\a$ points $S^{-t/2}x,\ldots, S^{t/2-1}x$.

The probability $\p(p)dp$ that $a_t\in (p,p+dp)$ can be computed from
\equ(3.13) with $T\gg t$ by using the Markov partition
$\EE_T$ constructed in \S3. The prisms of the partition $\EE_T$ are
naturally labeled by the allowed strings $\V\s\in\G_T$ (see the
paragraph preceding \equ(3.11)); for brevity we label them with a label
denoted $j$. If $E_j$ is a generic prism of this partition with axes
intersecting at $x_j$, consider the measure $\m_{T,t}$ defined by
setting, for every smooth $F$:

$$\ig_\CC F(x)\,\m_{T,t}(dx)=\fra{\sum_j\lis\L_{u,t}^{-1}(x_j)
F(x_j)}{\sum_j \lis\L_{u,t}^{-1}(x_j)}\Eq(4.6)$$

\0where $\lis\L_{u,t}(x)\=\prod_{j=-t/2}^{t/2-1} \L_u(S^jx)$ is defined
before \equ(3.4) and $x_j$ is a point in $E_j$.  In the limit $T\to\io$
the distribution $\m$ gives the correct probability to the values of an
observable $F$ which is smooth and has a $t$--dependence like the one in
\equ(4.4), (\ie like the one considered in theorem II, with
$f(x)\=\t_0^{-1}\ig_0^{t_0(x)} \a( S_\th x)d\th$) up to an error which
is a {\it factor bounded above and below uniformly in $t$}, by theorem
II.

The partition $\EE$ constructed in \S3 turns out to be time reversal
invariant, if the triangulation on $W^s$ is taken to be the $i$ image of
the triangulation on $W^u$: this means that if $E_j\in \EE$ then $i
E_j=E_{j'}\in\EE$ for some $j'$. Note that if we used another Markov
partition $\EE'$, constructed for instance via the classical proofs of
existence of Markov partitions of [S2],[Bo1], then the time reversal
symmetry of our system would still imply that $i\EE'$ is also a Markov
partition (as $W^u_x=i W^s_{ix}$) so that the partition obtained by
intersecting the two would still be a Markov partition, with the extra
property of being time reversal invariant; hence by using $\EE'\cap i
\EE'$ instead of $\EE$ we could still carry the argument that follows
which only depends on the time reversal invariance of the Markov
partition $\EE$ used for the construction of $\lis\m_+$ and not on the
particular Markov partition used. Note that if $\EE=i\EE$ then also the
partition $\EE_T$ in theorem II, $\EE_T=\cap_{q=-T}^T s^q\EE$, has
the property of time reversal invariance $\EE_T=i\EE_T$.

Thus, for instance, up to the mentioned error of a factor
$(T,t)$--independent:

$$\fra{\p(p)}{\p(-p)}=
\fra{\sum_{j,a_t(x_j)=p}\lis\L_{u,t}^{-1}(x_j)}
{\sum_{j, a_t(x_j)=-p}\lis\L_{u,t}^{-1}(x_j)}\Eq(4.7)$$

\0where the sums run over the labels $j$ of the
prisms $E_j\in \EE_T$ (with axes intersecting at $x_j$) verifying
$a_t(x_j)\in (p,p+\d)$.

By using the time reversal symmetry the above sums can be seen to
consist of sums with the same number of addends. This can be done by
pairing the contribution from $E_j\in \EE_T$ with that of $iE_j\=E_{j'}$
that can be easily seen to give, if $T$ is large (so that the prisms are
really small), a value $-p$ to $a_t(x_{j'})$ if $a_t(x_{j})=p$ and at
the same time$\lis\L_{u,t}(x_{j'})=\lis\L_{s,t}^{-1}(x_{j})$.
\footnote{${}^{14}$}{\nota
For instance, [ECM2], from the two identities $S^{-\t}(S^\t x)=x$ and
$S^{-\t}(i S^{-\t}x)=ix$ we deduce:

$$\eqalign{ &\dpr S^{\t}( S^{-\t}x)\,\dpr S^{-\t}(x)=1= \dpr S^{-\t}(
S^{\t}x)\,\dpr S^{\t}(x)\cr &\dpr S^{-\t}(i S^{-\t}x)\,i\,\dpr
S^{-\t}(x)=i
\cr}\eqno(*)$$

\0We see that, by applying the first relation with $x$ replaced by
$\tilde x'\=iS^{-\t}x$ and by applying successively the third (with
$x\to \tilde x'$) and the second relations (with $x\to S^{-\t}x$, using
also $x=i S^{-\t}\tilde x'$) in (*) one deduces:

$$\dpr S^\t(S^{-\t}\tilde x')i\dpr S^\t(S^{-\t}x)i=1\eqno(**)$$

\0Therefore: $\big(\det\dpr S^{\t}(S^{-\t}\tilde x')\big)^{-1} =\det \dpr
S^{\t}(S^{-\t}x)=e^{-(2N-1)\t \t_0\media\a a_\t(x)}$, where $\t_0$ is the
value of the timing interval in real units (for us $\t$ is an integer);
\ie, replacing $x$ with $S^{-\t/2}x$: $a_\t(x)=-a_\t(x')$.}
Hence the addends can be paired so that the ratios of corresponding
addends is:

$$\fra{\lis\L_{u,t}^{-1}(x_j)}{\lis\L_{s,t}(x_j)}= e^{(2N-1)\media{\a} p
t\,\t_0}\fra{b(S^{t/2}x_j)}{b(S^{-t/2}x_j)}\Eq(4.8)$$

\0where $\lis\L_{s,t}(x)$ is defined as
$\lis\L_{u,t}(x)\=\prod_{j=-t/2\t_0}^{t/2\t_0-1} \L_u(S^jx)$, with
$\L_c(x)$ being the absolute value of the determinant of the jacobian
matrix of $S$ as a map of $W^s_x$ into $W^s_{Sx}$. Eq. \equ(4.8) simply
follows from the fact that a volume element around $S^{-t/2}x$ varies
under the action of $S^t$ by a factor $\lis\L_{u,t}(x)\lis\L_{s,t}(x)
b(S^{t/2}x)/b(S^{-t/2}x)$ where $b(x)$ describes the transversality of
the intersection of the stable and unstable manifolds at $x$, as defined
in remark (iii) to definition 8: so that this equals $\exp{-(2N-1)t\,
\t_0\media{\a}_t(x)}$, (here $\t_0$ is considered extremely small, for
simplicity).

As repeatedly mentioned above \equ(4.6),\equ(4.7) require a correction:
but it has been remarked after \equ(4.6) that the correction is a factor
bounded by $e^{\pm B}$ for some $B$ which is $p$--independent because of
theorem II; and we can include in $B$ also the correction due to the
last ratio in \equ(4.8) (bounded uniformly in $t$, by transversality
property in the chaoticity assumption, see definition 8 in \S3).

This means that $\p(p)$ verifies ($N$ large):

$$\eqalign{ &\fra1{2 N t\,\t_0 p}\log
\fra{\p(p)}{\p(-p)}\tende{N\to\io,\,t\to\io} \media\a=p-{\rm
independent}\cr &\p(p)=e^{-\x(p)+p\media\a Nt\pm B}\cr}\Eq(4.9)$$

\0where $\x(p)$ is an even function of $p$ (over which we have no
control) and $B$ does not depend on $p$ nor on $t$ (but it may depend
on $N,E$).  The \equ(4.9) can be considered as a {\it large deviation}
result (both in $N$ and $t$): its peculiarity is the $p$ independence
of the coefficient of $p$ in the odd part othe argument of the
exponential, to leading order in $t$.

The $p,t,N$ independence of the first relation in \equ(4.9), and the
equality to $\media\a$ can be quite easily tested even if $N$ is
moderately large (but not too large: as in such case the measurements
simply could not be carried out): $N=64$ seems feasible.

The idea of the above test, and its realization in a model different
from \equ(4.1), is in the basic paper [ECM2] and it has been further
developed in [CG].  The test {\it does not} require the measurement of
the individual Lyapunov exponents: therefore it should be (relatively)
easy to carry out. Particularly after its feasibility has been proved
in similar models in [ECM2].

Finally it is remarkable that one can, at all, find tests of the
principle: in the case $E\ne0$ there is no really well established non
equilibrium thermodynamics with which one could compare the results of
the principle. Such general results would constitute the real test of
the principle from the point of view of Physics.  A non controversial
theory of non equilibrium thermodynamics could, actually, follow from
the principle.  For this, however, we must learn how to extract other
consequences, just as we learnt to extract consequences from the
Boltzmann Gibbs principle.  And it is important to stress that whatever
predictions it gives they should be as true as the statements derived in
equilibrium are: \ie essentially {\it exactly true}.  Therefore the
check above (and others that might be devised) will be satisfactory only
if it gives exactly the expected results, within the experimental
errors.  This seems to be the case in [ECM2] for the model considered
there, see also [CG].
\*

\0{\it\S5 Summary and outlook}
\numsec=5\numfor=1 \*

Here I regarded the Ruelle's principle as valid: to clarify its status I
considered of some interest setting up a list of assumptions that would
imply it rigorously. This essentially amounted to saying that the system
behaves as an Anosov system (or more generally I could have supposed
something like an axiom-A behaviour) and then use the theory of Sinai
(or, respectively, of Bowen and Ruelle) to "prove" the principle, see
\S3.  The interest of course is not in the mathematical theorem as I
just made enough assumptions to make it valid and checked its validity
by adapting the ideas of the key papers [S2],[Bo1],[R2], (while
attempting, to avoid repetitions, at describing a more intuitive
construction of Markov partitions), but rather the interest lies in the
clarification of the meaning of the property (C) in \S3.  One cannot be
too demanding on the matter of mathematical rigour: it should not be
forgotten that even the ergodic hypothesis of Boltzmann is far from
being proved, particularly in the generality one would want.

The "application" deviced in \S4 is not very satisfactory as a test of
validity of the principle because it can be performed only if $N,\t$
are not too large.  It would be, of course, nice to find a true
thermodynamic property that could be computed and tested via the
principle.

\01) One should remark that the above analysis should improve with
$N$, as the non chaotic phenomena should become less important:
they are excluded by the strict chaoticity assumption, see \S2, but
they are possibly present if the assumption is made in the loose form of
\S3, (C).  They would be certainly present if the hard core potential in
\equ(4.1) was replaced by a smooth finite range potential steeply
diverging at zero distance (by general results from KAM theory).

\02) Predictions analogous to the ones at the end of \S4 could be made
for other models, see [CG]: but except in the cases in [ECM2] the
numerical experiments do not seem to exist, yet.

\03) The new principle appears to play the role that the ergodic
hypothesis plays in equilibrium statistical mechanics: therefore one
may be led to think that the time of approach to equilibrium should be
of the order of the recurrence time on the attractor.  Having set up a
unified point of view for the equilibrium and the non equilibrium cases,
we can adopt the classical explanation of Boltzmann, [B96], [B02],
apparently relying also on an earlier suggestion by Thomson, [T],
intended to contradict such a hasty conclusion.  The rate of approach to
equilibrium is very short (essentially determined by the Boltzmann
equation in the case of rarefied gases, as exemplified, for instance, by
the Lorentz theory of conductivity in metals, [Be], closely related to
model 1) at least if one looks at the very few macroscopic observables
relevant for equilibrium thermodynamics and for the transport
coefficients: the reason being that such observables have the same value
on most of the surface $W^u_O$, see [Ga1].

\06) Concerning the particularity of the gaussian thermostat, which
could be called an ``unphysical fiction'', I think, see [Ga1],
that there should be, also in non equilibrium, several equivalent ways
of describing the same stationary distribution corresponding to
different $\lis \m$ and to different physical ways of reaching the
stationary state. And it might well be that the gaussian thermostat
turned out to be equivalent to other models of thermostats, which could
be described by rather different attractors.  For instance a stochastic
thermostat, in which a particle colliding with the wall comes out with a
maxwellian distribution at given temperature, will certainly be
described by a statistics $\lis \m$ which is absolutely continuous with
respect to the Liouville measure.\footnote{${}^{15}$}{\nota Note that a
stochastic model of thermostat is described by a stochastic differential
equation and therefore our discussion does not apply without some major
modification.} In the thermodynamic limit this might just be the same as
the result obtained with a statistics which, for finite $N$, is on a
strange attractor.  This mechanism is like the one realized by the
microcanonical and the canonical ensembles (the first is concentrated on
a set of configurations which has zero probability with respect to the
second, as long as $N<\io$).  This is clearly a question that requires
further investigations.

\vskip1.cm

\0{\it Aknowledgements:} I am indebted to J. Lebowitz, G. Eyink and Y.  Sinai
for many helpful comments. A special acknowledgement is for E.  Cohen
for stimulating my interest on the matter, and patiently explaining the
details of his results and his intuition of their relation with the
theory of chaotic systems: it was a great experience to profit of his
advice. I have included here many ideas that grew up from our
collaboration and that are included in our joint work [CG].  Support
from Ministero della Pubblica Istruzione, (grants named 40
\% and 60\%), and from CNR-GNFM for travel is acknowledged togheter with
support from the University of Granada, Spain, through the invitation
to deliver the present lectures.

\pagina

\*\*
\0{\it Appendix: Philosophical questions; and a few concrete ones}
\*
{\it The following is a guided series of problems to the general theory
of the Lyapunov exponents: it contains the ergodic commutative,
subadditive and non commutatitive theorems and the
Oseledec theorem.  The theorems are mostly philosophical, \ie they hold
with essentially no assumptions.  The problems are taken out of the
preprint of the book {\it Meccanica dei Fluidi}, in italian,
circulating in the form of a draft: the Oseledec theorem exposition is
essentially taken from [R4].}
\*
{\nota
{\bf(1)} If $\m$ is a probability measure on the Borel sets of $R^n$
and if $\D_n$ is a sequence of measurable sets such that $\sum_n
\m(\D_n)<+\io$ then almost all points are contained in at most a finite
number of sets in the sequence ({\it Borel Cantelli theorem}).  (Idea:
the set of the points in an infinite number of $\D_n$'s is
$N=\cap_{k=1}^\io(\cup_{h=k}^\io \D_h)$, of course.  Therefore
$\m(N)\le \sum_{h=k}^\io \m(\D_h)$ for all $k$'s, hence $\m(N)=0$
because the series converges.) \*

{\bf(2)} Let $(\CC,S,\m)$ be a triple formed by an invertible dynamical
system and an invariant probability measure on (the Borel sets of)
$\CC$ (\ie there is a zero measure set $N$ such that $S$ is invertible
outside $N$ and $\m(E)=\m(SE)=\m(S^{-1}E$) for all Borel sets $E\subset
\CC/N$).  Let $f$ be a (measurable) function bounded by a constant $K$
almost everywhere (with respect to $\m$).  Let $D_n$ be the set of
points $x$ such that some average of $f$ over a time $\le n$ is {\it
non negative}: \ie $m^{-1}\sum_{j=0}^{m-1} f(S^jx)\ge0$ for some $m\le
n$.  Then $\ig_{D_n} f(x)\m(dx)\ge0$.  ({\it Garsia maximal averages
theorem}).  (Idea: if $n=1$ the condition defining $D_1$ is simply
$f(x)\ge0$ and nothing has to be proved, besides the obvious.  If $n=2$
the condition defining $D_2$ is either $f(x)\ge0$ or $f(x)+f(Sx)\ge0$.
Therefore the {\it new} points, \ie those in $D_2$ but not in $D_1$, are
points $x$ where $f(x)<0$ which can be paired with point $Sx\in D_1$ so
that $f(x)+f(Sx)\ge0$.  Hence we can subdivide $D_2$ in the disjoint
union of $D_2/D_1\cup S(D_2/D_1)$ and of $D_1/S D_2$.  On the last set
it is $f(x)\ge0$ while the integral over the first union can be written
as $\ig_{D_2/D_1}(f(x)+f(Sx))\m(dx)$ because, by the invariance of $\m$
the integral $\ig_{S(D_2/D_1)} f(x)\m(dx)\=\ig_{D_2/D_1} f(Sx)\m(dx)$.
The case $n=3$ is only slightly more involved and it is left to the
reader (proceed in the "same way"), and, once understood, the general
case becomes crystal clear.) \*

{\bf (3)} The invertibility assumption
is not necessary in (2): the invariance in the ordinary sense
$\m(E)=\m(S^{-1}E)$ for all Borel sets $E$ is sufficient.  Prove this
statement.  (Idea: one has just to try to formulate what said by always
using $S^{-1}$: for instance $D_2$ will consist of the points in $D_1$
and the inverse images of those in $D_1$ such that
$f(x)+f(S^{-1}x)\ge0$, etc.) \*

{\bf(4)} Show that (2) implies the
almost everywhere existence of the limit $\lim_{k\to\io}\fra1k
\sum_{j=1}^{k-1} f(S^j x)$.  (Idea: let
$f_{sup}(x)=\limsup_{k\to\io}\fra1k \sum_{j=0}^{k-1} f(S^j x)$ and
$f_{inf}(x)=\liminf_{k\to\io} \fra1k \sum_{j=0}^{k-1}
f(S^j x)$.  The denial of the statement is that
$f_{sup}(x)>\a>\b>f_{inf}(x)$ on a set $D$ which has {\it non zero
measure} for a suitable pair $\a>\b$.  Quite absurd because on the
(obviously) invariant set $D$ the functions $f(x)-\a$ and $\b-f(x)$
would have some non negative average.  Hence by the previous theorem
their integrals over $D$ would have to be $\ge0$, but their sum would
therefore be $(\b-\a)\m(D)$ which would be $<0$, because
we are (foolishly) thinking that $\m(D)>0$!) \*

{\bf(5)} The boundedness assunption in (4) can be replaced by the
summability assumption $f\in L_1(\m)$. Furthermore if $\lis f(x)$
denotes the average of $f$, defined $\m$ almost everywhere, show that:
$$\eqalign{
&|\lis f|_{L_1}\=\ig_\CC \m(dx)\,|\lis f(x)|\le\ig_\CC \m(dx)|f(x)|\=
|f|_{L_1}\cr
&\ig_\CC \lis f(x)\m(dx)=\ig_\CC f(x) \m(dx)\cr}$$
If there are no non trivial functions $f(x)$ which are constant of
motion $\m$ almost everywhere, \ie if the system is {\it ergodic}, then
the function $\lis f(x)$ is a constant (almost everywhere) and
$\lis f=\ig_\CC\m(dy) f(y)$ for all $f\in L_1(\m)$.
\*

{\bf(6)} Let $(\CC,S,\m)$ be a dynamical system as in problem (4). And
let $f_n(x)$ be a sequence of measurable functions such that:
$$|f_1(x)|<K,\qquad f_{n+m}(x)\le f_n(x)+f_m(S^nx)\qquad\hbox{$\m$
almost everywhere}$$
Suppose $\m$ ergodic; by applying the ergodic theorem (3),(4) above
show the existence, $\m$ almost everywhere, of the limit $\lim_{n\to\io}
\fra1n f_n(x)=\lis f(x)$, ({\it Kingman's subadditive ergodic theorem}).
(Idea: Remark that the functions:
$$f_{sup}(x)=\limsup_{n\to\io} \fra1n f_n(x),\qquad
f_{inf}(x)=\liminf_{n\to\io} \fra1n f_n(x)$$
verify $f_{sup}(Sx)\ge f_{sup}(x)$ and $f_{inf}(Sx)\ge f_{inf}(x)$
(because $f_n(x)\le f_1(x)+f_{n-1}(Sx)$, then divide by $n$), so that
the $\m$ invariance of $\m$ implies that they are constants of motion
(because $\ig (f_{sup}(Sx)-f(x))d\m=0$).  Therefore, $\m$ almost
everywhere, $f_{sup}(x)=\b$ and $f_{inf}(x)=\a$ where $\a<\b$ are
suitable constants.  Let us consider a number $\h>0$ such that
$\a+\h<\b$.\\ Let $\D_n$ be the set of points where $f_n(x)\le\a+\h$
for at least one value of $m\le n$: hence $\lim_{n\to\io}\m(\D_n)=1$.
Given $\e>0$ there is, therefore, a $n_\e$ such that
$\m(\D_{n_\e}^c)<\e$ if ${}^c$ denotes the complementation operation on
sets.  If $x\in\CC$ one can suppose that the frequency of visit of $x$
to $\D_{n_\e}^c$ is $<\e$, because such frequency is the average value
of $\chi_{\D^c_{n_\e}}(S^jx)$ (over $j$) if $\chi_\D$ denotes the
characteristic function of the set $\D$ (by the ergodicity).\\ Consider
the sequence of times $j_1<j_2<\ldots$ when, instead, $S^{j_k}x\in
\D^c_{n_\e}$: it follows that the number $p$ of such $j$'s with $j_k\le
T$ is such that $p/T<\e$ for $T$ large enough (as $p/T$ tends to the
frequency of visit to $\D^c_{n_\e}$).\\ Let $k_0$ be the first time
$\le T$ in which $S^{k_0}x\in \D_{n_\e}$ and let $k_0'$ be the largest
integer $\le T$ such that $f_{k'_0-k_0}(S^{k_0}x)\le (k'_0-k_0)(\a+\h)$.
The point $k'_0+1$ must be one of the $j$'s, otherwise it could not be
the largest integer $k'$ such that $\fra1{k'-k_0}
f_{k'-k_0}(S^{k_0}x)\le \a+\h$ by the subadditivity of $f$, unless of
course $k'_0=T$.\\ Let $k_1>k'_0$ be the first value, with $k_1\le T$,
not among the $j$'s, \ie such that $S^{k_1}x\in\D_{n_\e}$ and let
$k'_1>k_1$ be the largest value such that $f_{k'_1-k_1}(S^{k_1}x)\le
(k'_1-k_1) (\a+\h)$, and so on.\\ In this way a sequence
$[k_0,k'_0],\ldots,[k_s,k'_s]$ is constructed in the interval $[0,T]$.
All the values $k<k'_s$ outside the intervals must be among the $j$'s
(hence their number is $\le p$).  The last value $k'_s$ will be,
possibly, followed by a string of values among the $j$'s but the first
value $k_{s+1}$ that does not have this property, if existing at all,
{\it must} be within $n_\e$ of the value $T$, otherwise we could form
the interval $[k_{s+1},k'_{s+1}]$.  Therefore the number of values
outside the intervals is bounded above by $p+n_\e$.  The subadditivity
then implies:
$$\eqalign{\txt
&\fra1T f_T(x)\le \fra1T \big(K(p+n_\e)+\sum_{i=1}^s
f_{k'_i-k_i}(S^{k_i}x)\big)\le\cr
\txt&\le \fra1T\big(K(p+n_\e)+
T(\a+\h)\big)\tende{T\to\io}\a+\h\cr}$$
which shows the one {\it cannot} contemplate the case $\a<\b$, by the
contradiction it does provoke: hence $\a=\b$.)
\*

{\bf(7)} Show that (6) implies that $\lim_{n\to\io}\fra1n(x)=\lis f(x)$
is a $\m$-almost everywhere constant, and:
$$\lim_{n\to\io}\fra1n f_n(x)\=\lis f=\lim_{n\to\io}\fra1n\ig_\CC
f_n(x)\m(dx)=\inf_n \fra1n\ig_\CC f_n(x)\m(dx)$$
(Idea: subaddititivity implies $f_n(x)\le f_1(x)+f_{n-1}(Sx)$; hence
$\lis f(x)\le \lis f(Sx)$ and $0\le\ig_\CC (\lis f(Sx)-\lis f(x))d\m=0$
imply $\lis f(Sx)=\lis f(x)$ $\m$--almost everywhere. Then ergodicity yields
that $\lis f$ is constant.  By dominated convergence the first limit
relation follows.  The sequence $n\to \media{f_n}\=\ig_\CC
f_n(x)\m(dx)$ is subadditive and bounded by $K$; hence $\fra1n
\media{f_n}\tende{n\to\io}\inf_n \fra1n\media{f_n}$, by an elementary
argument.) \*

{\bf(8)} Show that the assumption $|f_1(x)|<K$ in (6) can be replaced
by the summability of $f^+$, $f^+(x)\in L_1(\m)$, if $f^+(x)=\max(0,
f_1(x))$. (Idea: no idea is necessary; just a careful examination of the
proofs in (6),(7).)
\*

{\bf(9)} Show that the ergodic theorem implies that if $N$ has zero
measure and $S^jN$ is measurable then $\m(SN)=0$. (Idea the frequency of
visit of the motion starting at $x$ to the set $SN$, $\f_x(S N)$, is
equal to that to $N$ itself: $\f_x(N)=\f_x(SN)$; hence by the ergodic
theorem $\m(SN)=\ig \f_x(N)d\m\=\ig \f_x(SN)d\m=\m(N)=0$.)
\*

{\bf(10)} Define, via the ergodic theorem, the "future" and "past"
averages of $f\in L_1(\m)$ as the limits
$f^\pm(x)=\lim\fra1n\sum_{j=0}^{n-1}
f(S^{\pm j}x)$. Show that $f^+(x)=f^-(x)$ $\m$--almost everywhere.
(Idea: Let $D,\a,\b$ be such that $\m(D)>0$ and $f^+(x)>\b>\a> f^-(x)$
for $x\in D$. Let $D^+_n$ be the set of points $x\in D$ such that
$\fra1m \sum_{j=0}^{n-1}f(S^jx)\=\media{f}^+_m(x)>\b$ for all $m\ge n$.
Let $D^-_n$ be the correspoonding set where $\media{f}^-_m(x)<\a$. Then,
if $n$ is large enough $S^{-(n-1)}D^-_n\cap D^+_n\ne \emptyset$, because
the quantities $\m(D^-_n)\=\m(S^{-(n-1)}D^-_n)$ and $\m(D^+_n)$ are both
very close to $\m(D)$ for $n$ large. Hence for $x$ in the latter
intersection it is $\a>\media{f}^-_n(x)\=\media{f}^+_n(S^{-(n-1)}x)>\b$
by the properties defining  $D,D^+_n,D^-_n$. This is impossible.)
}
\*
The guided sequence of problems will now come "closer to Earth" by
studying some finite matrix theorems.
\*
{\nota

{\bf(11)} Let $L$ be a real $d\times d$ non singular matrix. Consider
the matrix $M_n\= ((L^*)^nL^n)^{1/2}$ and call $t^{(n)}_j$ the
eigenvalues of $M_n$ ordered by decreasing size. Show that the largest
eigenvalue of $M_n$ is such that the limit $\fra1n\log
t^{(n)}_1\tende{n\to\io}\l_1$
exists. Here ${}^*$ means transposition with respect to the scalar
product $(u,v)=\sum_i u_i v_i$.
(Idea: first show that the eigenvalues $t_j^{(n)}$ of
$((L^T)^nL^n)^{1/2n}$, ordered according to decreasing size,
do have a limit as $n\to\io$.
Note that $f_n=\log |L^n|$ verifies subadditivity $f_{n+m}\le f_n+f_m$
and $f_n\le n \log|L|$, if $|L|$ is the norm of the matrix $L$
(\ie $|L|=\max |Lv|/|v|$, with $|v|=(v,v)^{1/2}$). Hence the
limit $n^{-1}\log |L^n|\tende{n\to\io}t_1=\inf_n\fra1n\log |L^n|$
exists, and $\fra1n f_n\ge \min \log |\m_j|$ if $\m_j$ are the eigenvalues of
$L$. But $|L^n v|=(L^n v, L^n v)^{1/2}=(v,(L^*)^nL^n v)^{1/2}\le
\max_{1\le j\le d}(t^{(n)}_j)^{1/n}$, if $|v|=1$. Hence
the largest eigenvalues have the appropriate convergence property.)
\*

{\bf(12)} Think of the vectors in $R^d$ as functions $i\to u_i$ on the
finite space $F\=(1,2,\ldots,d)$. Let $(R^d)^{\wedge q}$ be the space of the
functions on  $F^q$ which are {\it antisymmetric}: these are the
functions $u_{i-1\ldots i_q}$ which are antisymmetric. Define a
scalar product on such functions by setting $(U,v)\=\sum u_{i_1\ldots
i_q}v_{i_1\ldots i_q}$. Define the matrix $L^{\wedge q}$ by setting:
$$(L^{\wedge q} u)_{i_1\ldots i_q}=\sum_{j_1\ldots j_q}^{1,d}
L_{i-1j_1}L_{i_2j_2}\ldots L_{i_q}{j_q} u_{j_1\ldots j_q}$$
Show that the result of (12) implies that the matrix
$(((L^{\wedge q})^*)^n(L^{\wedge q})^n)^{1/2n}$, where the ${}^*$
denotes the adjunction operation with respect to the scalar product
$(u,v)$, is such that its largest
eigenvalue logarithm divided by $n$ has a limit as $n\to\io$.

\*

{\bf(13)} Let $(t^{(n)}_j)^2$ be the eigenvalues of $(L^*)^nL^n$ in
decreasing order, repeated according to multiplicity.
Prove that the eigenvalues of $((L^{\wedge q})^n)^*
(L^{\wedge q})^n$ are just the products of the $q$-ples
of eigenvalues $(t^{(n)}_{j_1})^2\ldots
(t^{(n)}_{j_q})^2$ with $j_i\ne j_j$ (\ie products of $q$--ples of pairwise
distinct eigenvalues). (Idea: no idea is necessary.)
\*

{\bf(14)} Combine (12) and (13) to infer that the limits $\fra1n\log
t^{(n)}_j=\l_j$ exist for all $j=1,\ldots,d$. Let $\lis\l_1,\ldots,\lis
\l_s$ be the {\it distinct} limits $\l_j$ and let $m_1,\ldots,m_s$
be their multiplicities (\ie the number of $\l_i$ which are equal to
$\lis \l_j$). Verify that $\sum_{i=1}^s m_i=d$ and define $r(i)=j$ if
$\l_i=\lis\l_j$.
\*

{\bf(15)} Let $U^{(n)}_1$ be the linear space spanned by the
first $m_1$ eigenvectors of $\L_n=((L^*)^nL^n)^{1/2}$; likewise
$U^{(n)}_2$ will be the space spanned by the next $m_2$, and so on
until $U^{(n)}_s$ is defined.  Show that the notion of multiplicity
introduced in (14) is even more justified by proving the following
"orthogonality" property between unit vectors $u\in U^{(n)}_r$ and $u'\in
U^{(n+k)}_{r'}$, $k\ge0$. Given $\d>0$ there exists $C>0$ such that:
$$|(u,u')|\le C e^{-(|\lis\l_r-\lis\l_{r'}|-\d)n}$$
(Idea: nothing to prove if $r=r'$, of course. The "easy case" is $r'>r$.
Let $\L_{n+1}=\sum_{i=1}^d t^{(n+1)}_i P_i$ be a spectral decompositon
for the matrix $\L_{n+1}$.  Then supposing $n$ so large that for $m\ge
n$ it is $|\fra1m\log t^{(m)}_j-\lis\l_i|<\d_1$ for all $r(j)=i$ and all
$i$, where for a given $\d_1>0$, $\d_1<\fra12\min(|\l_r-\l_{r'}|)$:
$$\eqalign{\txt
|(u,u')|\=&\txt e^{-(n+1)\lis\l_{r'}}
\max_{u'\in U^{(n+1)}_{r'}}|(u,\sum_{r(i)=r'}
e^{(n+1)\lis\l_{r'}} P_i u')\=\cr\txt
\=&\txt e^{-(n+1)\lis\l_{r'}}\max_{u'\in U^{(n+1)}_{r'}}|(\sum_{r(i)=r'}
e^{(n+1)\lis\l_{r'}} P_i  u,u')\le\cr\txt
\le&\txt  e^{-(n+1)\lis\l_{r'}}\big|\sum_{r(i)=r'}
e^{(n+1)\lis\l_{r'}} P_i u\big|\le\cr\txt
\le&\txt
e^{-(n+1)\lis\l_{r'}+(n+1)\d_1}\big|\sum_{r(i)=r'}
t^{(n+1)}_i P_i u\big|\le\cr\txt
\le&\txt
e^{-(n+1)\lis\l_{r'}+(n+1)\d_1}\big|((L^*)^{n+1}
L^{n+1})^{1/2}u\big|\cr\txt
=&\txt  e^{-(n+1)\lis\l_{r'}+(n+1)\d_1}(u,
(L^{n+1})^*L^{n+1}u)^{1/2}\cr}$$
Thus, again by the spectral theorem and by $|T T'|\le |T|\,|T'|$ we deduce:
$$\eqalign{|(u,u')|\le &
e^{-(n+1)(\lis\l_{r'}-\d_1)}\big|L^{n+1}u\big|
\le\cr\le & e^{-(n+1)(\lis\l_{r'}-\d_1)}|L|\big|L^{n}u\big|
\le e^{-(n+1)(\lis\l_{r'}-\lis\l_r-2\d_1)}|L|\cr}$$
because $u\in U^{(n)}_r$. This completes the proof in the case $r'>r$
and $k=1$. The case $k>1$ is simply obtained from the case $k=1$ by
applying the latter inequality $k$ times. One finds, since the
series converges, the result with a $C$ that can be taken
$C_1=|L|(1-e^{-x})^{-1}$ if
$0<x<\min(|\m_r-\m_{r'}|)-2\d_1$ for all $n$ large enough and all
$k\ge0$.\\
Consider the case $r>r'$. Let $u_\a$ be an orthonormal base with the
first $m_1$ vectors spanning $U^{(n)}_1$, the next $m_2$ spanning
$U^{(n)}_2$ and so on; let $u'_\a$ be the corresponding base for $n+k$.
The orthogonal matrix $W_{\a\a'}=(u_\a,u'_{\a'})$ verifies the
inequalities:
$$\cases{
|W_{\a,\a'}|\le 1&for all $\a,\a'$\cr
|W_{\a,\a'}|\le
C_1 e^{-|\lis\m_{r(\a)}\lis\m_{r(\a')}-2\d_1| n}&
if $r(\a')>r(\a)$\cr}$$
If $r(\a')<r(\a)$, {\it however}, the orthogonality implies that
$W_{\a\a'}\=(W^{-1})_{\a'\a}$. The latter quantity is "just" the
determinant obtained by deleting the row $\a$ and the column $\a'$
from the matrix $W$. Such determinant consists in a sum of $(d-1)!$
products of $d-1$ matrix elements $W_{ij}$ picked up in pairwise
distinct rows and columns (by Cramers' rule). Since we are interested in
the non diagonal elements of $W_{\a\a'}$ we see that in each product
there must be at least enough factors $W_{\b\b'}$ with $r(\b')>r(\b)$
so that $\sum \m_{r(\b')}-\m_{r(\b)}\ge \m_{r(\a)}-\m_{r(\a')}$ (hint:
check this first when there are no degenracies, \ie $r(\a)\=\a$). Hence
we use for such factors the second inequality and just bound the others
by $1$. We finally get the result with $C=C_1(d-1)!$ if $2\d_1
(d-1)<\d$.)
\*

{\bf(16)} Show that (15) implies that the planes $U^{(n)}_j\subset R^d$
have a limit $U_j$ as $n\to\io$ and the planes $U_j$ are pairwise
orthogonal. (Idea the planes $U^{(n+h)}_1$ and
$U^{(n+k)}_1$ must form with the planes $\bigoplus_{r>1} U^{(n)}_r$ an
angle closer to $90^o$ by an amonut prefixed arbitrarily if $n$ is large
enough. Hence they form a Cauchy sequence of planes converging to some
$U_1$. The other planes are treated analogously.)
\*

{\bf(17)} Check that the previous problems (11)$\div$(16) imply
that if $L$ is a real $d\times d$ non singular matrix the limit
$\lim_{n\to\io} ((L^*)^nL^n)^{1/2n}=D$ exists and it is a non singular
positive definite matrix, with eigenvalues equal to the absolute values
of those of $L$, counted according to multiplicity.  The eigenspaces
$U^{(n)}_j$ of $D_n\= ((L^*)^nL^n)^{1/2n}$ spanned by the $r(j)$
eigenvalues of $D_n$ that converge to the $j$-th distinct limit value
are planes that converge to limit planes $U_j$ which are the
eigenplanes of $D$ which correspond to distinct eigenvalues.  (Idea the
only thing still to check is the relation between the eigenvalues of
$D$ and those of $L$.  Suppose for simplicity that the eigenvalues of
$L$ have pairwise distinct absolute values (hence they must be all
real) and order them by decreasing absolute values $\m_1,\ldots,\m_d$.
Let $L=\sum_j \m_j v_j\times v^*_j$ be the spectral resolution of $L$
where $L v_j=\m_j v_j$, $L^* v^*_j=\m_j v^*_j$ and $(v^*,v)=1$.  Clearly
$|L^n v_j|=|\m_j|^n |v_j|$.  Hence $|\m_1|=\l_1$.  The other
equalities can be deduced for instance by the "trick" of considering
the matrices $L^{\wedge q}$, as in (13).)\vfill}

{\it Going back to the realm of abstract thinking the following guided
problems, combining the concrete theory (11)$\div$(17) and the
philosophical results (1)$\div$(10), lead to the Oseledec theorem.}
\vfill
{\nota {\bf(18)} Given an ergodic dynamical system $(\CC,S,\m)$ with an
invariant distribution $\m$ such that there is a zero $\m$ measure
invariant set $N$ outside which the transformation $S$ is
invertible and non singular, consider the matrix $\dpr S^n(x)\=T^n(x)$.
Suppose that $|\det T(x)|\ge\e>0$ and that $|T(x)|<E$ in $\CC/N$.
Check that $T_n(x)=T(S^{n-1}x)\cdot\ldots\cdot T(Sx)\cdot T(x)$,
and that $f_n(x)=\log| T_n(x)|$ verifies the subadditivity property
of the subadditive ergodic theorem of (6). Therefore:
$\lim_{n\to\io} \fra1n \log|T_n(x)|=\l_1(x)$ exists $\m$ almost
everywhere.
\*

{(\bf(19)} Define the matrices $T_n^{\wedge q}(x)$ on $R^{\wedge q}$ as
in (12),(13) above, and by repeating the argument there (with (18)
replacing (11)) prove that the limits $\lim_{n\to\io}\fra1n\log
t^{(n)}_j(x)=\l_j(x)$ exist almost everywhere for $\m$ almost all $x$.
\*

{\bf(20)} Show that the analysis in (15) can be repeated word by word
even when the matrices $L^n$ are replaced by $T^n(x)$, in the points
where the limits in (19) exist, \ie almost everywhere. The matrix
$D=\lim_{k\to\io}((T^*_k(x)T_k(x))^{1/2k}$ has eigenplanes
$U_1(x),\ldots,U_{s(x)}(x)$. Show that the spaces
$V_j(x)=U_{j}(x)\oplus\ldots\oplus U_{s(x)}$ can be identified with the
system of scaling planes for $S$ of definition 6, \S2.
\*

{\bf(21)} The contraction exponents $\l_j(x)$ are defined almost everywhere
and they are constants of motion togheter with their multiplicities
$m_j(x)$. Therefore they are $\m$ almost everywhere constants. (Idea: By
(20) and an argument like that in (17).)
\*

{\bf(22)} Check that all the above results can be derived by just
requiring that $T(x)$ is such that $\log^+|T(x)|$, where $\log^+$
denotes the positive part of the logarithm function, is $\m$--summable.
(Idea: just careful examination of the proofs.) \*

{\bf(23)} If $T(x)$ is replaced by any $d'\times d'$--matrix valued
function $x\to O(x)$ with $\log^+|O(x)|$ $\m$--summable the "same
results" can be proved.  For instance if
$O_n(x)=O(S^{n-1}x)\cdot\ldots\cdot O(x)$ then
$D_n=(O_n^*(x)O_n(x))^{1/2n}$ has a limit $D$ as $n\to\io$ which is
almost surely independent of $x$, and the eigenspaces, spanned by the
eigenvalues of $D_n$ whose eigenvalues converge to the same limit
$\lis\l_j$, converge to the eigenspace of $D$ with eigenvalue $\l_j$.
In other words the above theory extends trivially to a theory of
products of random matrices $O(x)$ generated by selecting randomly by a
distribution $\m$ on $\CC$ a point $x$ and the corresponding random
matrix $O(x)$ and by multipliyng $O(S^{n-1}(x)\ldots O(x)$.  \*

{\bf(24)} Let $\tilde T(x)=\dpr S^{-1}(x)$, $\dpr S(x)=T(x)$.  Then
$\tilde T(x)=T(S^{-1}x)^{-1}$ and $\tilde T_k(S^kx)= \dpr S^{-k}(S^kx)$
can be written either $\tilde T(Sx)\ldots \tilde T(S^{k}x)$ or also as
$T^{-1}(x)\ldots T^{-1}(S^{k-1}x)$.  Show that if $x$ admits a system of
contracting planes, \ie $\m$--almost
everywhere, the eigenvalues $\tilde t^{(k)}_j$ (ordered by decreasing
size) of $(\tilde T^*_k(S^kx)\tilde T_k(S^kx))^{1/2}\=\tilde D_k$ have
the limit: $\lim\fra1k \log \tilde t^{(k)}_j=\tilde\l_j=-\l_{d-j}$,
$\m$--almost everywhere.
(Idea: note the following expression for $\tilde T_k$:
$\tilde T_k^*(S^kx)\=\tilde T^*(S^kx)\ldots \tilde
T(S x)$ so that the matrices are multiplied in the correct order for
the application of (23): hence $(\tilde T_k(S^kx)\tilde
T^*_k(S^kx))^{1/2k}$ (identical to $(T_k^*(x) T_k(x))^{-1}$)
converges to a matrix $\tilde D\=D^{-1}$, $\m$--almost
surely (or if $x$ admits a contracting system of planes).
Thus the limits exist, because the spectrum of $T^*T$ and that
of $T T^*$ coincide in general, if $T$ is non singular; their relation
with the opposites of the contraction exponents $\l_j$ in the forward
direction is derived from $\tilde T_k(S^kx)\tilde
T^*_k(S^kx)=(T_k^*(Sx) T_k(Sx))^{-1}$.) \*

{\bf(25)} In the context of the problems (20),(24) consider the system
of planes $\hat V_j(x)$, at $x$, defined by the eigenplanes $U_j(x)$
of the matrix $D$ of problem (20) by $\hat V_j=
U_{s-j+1}\oplus\ldots\oplus U_1(x)$, and existing $\m$--almost
everywhere.  Show that if $w_k\in U_j(S^kx)/U_{j+1}(S^kx)$ and
$w_k=\dpr S^k(x)u$ (hence $u\in \hat U_j(x)/\hat U_{j+1}(x)$) then
$\lim_{k\to\io}\fra1k\log |\tilde T(S^k)w_k|/|w_k|=-\lis \l_{s-j}$.
Why one should not call $\hat V_j(x)$ an expanding plane when $j$ is
such that $\lis\l_{s-j'}<0$  for all $j'<j$ and positive otherwise?
\*

{\bf(26)} Let us call the spectrum of a sequence of matrices $O_n$
the $d$ numbers $\o_1,\ldots,\o_d$ obtained by considering the logarithms
of the eigenvalues
of $(O^*_kO_k)^{1/2k}$, $o^{(k)}_1\ge\ldots o^{(k)}_d$, and setting,
when they exist: $\lim_{k\to\io} o^{(k)}_j=\o_j$. Show that the spectrum
of the sequence $O_k(x)=T_k(S^{-k}x)$ coincides with that of $T_k(x)$
$\m$--almost surely.  (Idea: if the functions $n\to f_n(x)$ and $n\to
f_n(S^nx)$ are subadditive, in the sense of problem (6), the functions
$n\to f_n(S^nx)$ and $n\to f_n(x)$ are such that $F=\lim_{n\to\io}
\fra1n f_n(S^nx)$ and $G=\lim_{n\to\io}\fra1n f_n(x)$ exist and are
constant $\m$--almost everywhere, by (6) above.  Suppose that $F>G$.
Consider the sets $D_n$ and $D'_n$ consisting in the points where
$\fra1m f_m(S^mx)> F-\e>G$ for all $m\ge n$, or respectively where
$\fra1m f_m(x)<F-\e$ for {\it all} $m\ge n$.  Then $\m(D_n),\m(D'_n)\to
1$.  Hence for $n$ large enough $S^{-n}D_n'\cap D_n\ne \emptyset$: if
$x$ is in this set then $\fra1n f(S^nx)<F-\e$ because $S^nx\in D_n'$
and $\fra1n f_n(S^nx)>F-\e$ because $x\in D_n$.  Hence this
contradiction shows that it is sufficient to check that the two
functions are subadditive: this comes from the inequality
$|AB|\le|A||B|$.)
\*

{\bf(27)} The spectrum, see (26), of $\tilde T_{-k}(x)\=\tilde
T(S^{-(k-1)}x)\ldots \tilde T(x)$ is the opposite of that of $T_k(x)$:
$\o_j=-\lis\l_{d-j+1}$.  (Idea: $\tilde
T_{-k}(x)=(T_k(S^{-k}x))^{-1}$; hence $\tilde T_{-k}(x)^*\tilde
T_{-k}(x)=((T_k(S^{-k}x)(T_k(S^{-k}x))^*)^{-1}$; but the spectrum of
the last matrix is the same as that of the matrix
$(T_k(S^{-k}x)^*T_k(S^{-k}x))^{-1}$ which by (26) leads to the result.)
\*

{\bf(28)} The forward and backward systems of scaling
planes exist $\m$--almost
everywhere and have opposite corresponding exponents:
$\l_j=-\o_{d-j+1}$.  (Idea: this is a corollary of what already
discussed in the few preceding problems.)
\*

{\bf(29)} Show that if $V_{-j}(x)$ are the
contracting planes for $S^{-1}$, defined $\m$--almost
everywhere, then $V_r(x)\cap V_{-(s-r+2)}(x)=0$ and $V_r(x)\oplus
V_{-(s-r+2)}=R^d$,
$\m$--almost everywhere.
(Idea:  since, by definition, the dimensions of $V_r$ and
of $V_{-(s-r+2)}$
are complementary to $d$ one only has to check the first statement.
Note that if $u\in V_r(x)$ it is $|T_k(x)u|\le e^{k(\l_r+\d)}|u|$ for any
prefixed $\d>0$ and for $k$ large enough and all $x\in D_k$ where $D_k$
is a set such that $\m(D_k)\to1$. For the same reason $|T_{-k}(S^kx)v|\le
e^{-(\l_{r-1}+\d)k}|v|$ for $v\in V_{-r}(S^kx)$ provided $S^kx\in D_k$.
Since $T_{-k}(S^kx) V_{-(s-r+2)}(S^kx)=V_{-(s-r+2)}(x)$ we see that all
 the $u\in
V_{-(s-r+2)}(x)$ have the form $u=T_{-k}(S^kx)v$ with $v\in V_{-r}(S^kx)$.
So that using $T_k(x)T_{-k}(S^kx)=1$ we get:
$$\eqalign{
&|T_{-k}(S^kx)v|\le e^{-(\l_{r-1}-\d)k}|T_k(x)T_{-k}(S^kx)v|\to\cr
&\to |T_k(x)u|\ge e^{(\l_{r-1}-\d)k}|u|\qquad {\rm for}\ x\in
S^{-k}D_k,\ u\in V_{-(s-r+2)}(x)\cr}$$
and if $x\in S^{-k}D_k\cap  D_k$ this is impossible because
$\l_{r-1}>\l_r$ if also $u\in V_r(x)$ (so that $|T_k(x)u|\le
e^{(\l_r-\d)k}|u|$) and $\d$ is smaller than the twice the minimal
difference between the scaling exponents.)
\*

{\bf(30)} The planes $W_j(x)=V_{j}(x)\cap V_{-(s-j+1)}(x)$ are such that
$$\fra1k \log|T_{\pm k}(x)u|\tende{k\to\io}\l_j\qquad{\rm for}\ u\in
W_j(x)$$
and $V_j=W_j\oplus\ldots\oplus W_{s}$, $V_{-(s-j+2)}=W_1\oplus\ldots\oplus
W_{-(j-1)}$. Also $T(x)W_j(x)=W_j(Sx)$. This also implies the relation
between the dynamical bases claimed in the theorem II. (Idea: this is
just a corollary, or a summary, of the previous problems.)
\*

{\bf(31)} Show that if $\m$ is non invariant the foward and backward
Lyapunov exponents (and the relative systems of planes) may exist and
be different.  (Idea: consider the example on $\CC=[-1,1]\times T^2$,
where $T^2$ is the two dimensional torus.  Let $x=(z,\f_1,\f_2)$ be a
point in $\CC$.  Let $z\to f(z)$ be a smooth map such that
$f^n(z)\tende{n\to\pm\io}\pm1$, and let $\n(z)=2$ if $z>0$ and
$\n(z)=1$ if $z<0$; define:
$$x'=(z',\f_1'\f_2')=\cases{f(z)\cr \f_1+\n(z)\f_2\quad\hbox{mod}\,2\p\cr
\n(z)\f_1+(\n(z)^2+1)\f_2\quad\hbox{mod}\,2\p\cr}$$
and check the statements using $\m=dz\,d\f_1\,d\f_2/2(2\p)^2$.)
\*

{\bf(32)} Find the positive and negative Lyapunov exponents in the
example suggested in problem (31); find also the dynamical bases (of the
points which admit them).
}
\*\*
{\bf References.}
\*
\item{[B96]} Boltzmann, L.: {\it Entgegnung auf die
w\"armetheoretischen Betrachtungen des Hrn.  E.  Zermelo}, english
translation in S.  Brush, "Kinetic Theory", {vol. 2}, 218--, Pergamon
Press.

\item{[B02]} Boltzmann, L.: {\it Lectures on gas theory}, english
edition annotated by S.  Brush, University of California Press,
Berkeley, 1964.

\item{[Bo1]} Bowen, R.: {\it Equilibrium states and the ergodic theory of
Anosov diffeomorphisms}, Lecture notes in mathematics, vol. {\bf 470},
Springer Verlag, 1975. See also [Ga3].

\item{[Bo2]} Bowen, R.: {\it Markov partitions are not smooth},
Proceedings of the American Mathematical Society, {\bf 71}, 130--132,
1978.

\item{[BSC]} Bunimovitch, L., Sinai, Y., Chernov, N.I.: {\it Statistical
properties of two dimensional hyperbolic billiards}, Russian
Mathematical Surveys, {\bf 45}, n. 3, 105--152, 1990.

\item{[CB]} Cohen, E.G.D., Berlin, T.H.: {\it Note on the derivation of
the Boltzmann equation from the Liouville equation}, Physica, {\bf 26},
717--729, 1960.

\item{[CELS]} Chernov, N.I., Eyink, G.L., Lebowitz, J.L., Sinai, Y.:
{\it Steady state electric conductivity in the periodic Lorentz gas},
Communications in Mathematical Physics, {\bf 154}, 569--601, 1993.

\item{[CG]} Cohen, E., Gallavotti, G.: {Non equilibrium ensembles},
preprint 1994, in preparation.

\item{[Dr]} Dressler, U.: {\it Symmetry property of the Lyapunov
exponents of a class of dissipative dynamical systems with viscous
damping}, Physical Review, {\bf 38A}, 2103--2109, 1988.

\item{[ECM1]} Evans, D.J., Cohen, E.G.D., Morriss, G.P.: {\it Viscosity of a
simple fluid from its maximal Lyapunov exponents}, Physical Review, {\bf
42A}, 5990--\-5997, 1990.

\item{[ECM2]} Evans, D.J., Cohen, E.G.D., Morriss, G.P.:
{\it Probability of second law violations in shearing steady flows},
Physical Review Letters, {\bf 71}, 2401--2404, 1993.

\item{[FZ]} Franceshini, V., Zironi, F: {\it On constructing Markov
partitions by computer}, Journal of Statistical Physics, {\bf 40},
69--91, 1985.

\item{[Ga1]} Gallavotti, G.: {\it Ergodicity, ensembles, irreversibility
in Boltzmann and beyond}, preprint Physics department, University of
Wien, deposited in $mp\_ arc@math.utexas.edu$, \#94-66.

\item{[Ga2]} Gallavotti, G.: {\it Ising model and Bernoulli shifts}:
Communications in Mathematical Physics, {\bf 32}, 183--190, 1973.

\item{[Ga3]} G. Gallavotti, {\it Aspetti della teoria ergodica, qualitativa
e statistica del moto}, Quaderni dell' UMI (Unione Matematica Italiana),
vol. 21, Bologna, 1981.

\item{[GG]} Garrido, P., Gallavotti, G.: {\it Billiards correlation
functions}, in {\sl mp${}\_$arc@math.utexas.edu}, \#93-279, to
appear in Journal of Statistical Physics, 1994.

\item{[LPR]} Livi, R., Politi, A., Ruffo, S.: {\it Distribution of
characteristic exponents in the thermodynamic limit}, Journal of
Physics, {\bf 19A}, 2033--2040, 1986.

\item{[P]} Pesin, Y.: {\it Dynamical systems with generalized hyperbolic
attractors: hyperbolic, ergodic and topological properties}, Ergodic
Theory and Dynamical Systems, {\bf 12}, pp.123-151, 1992.

\item{[PH]} Posch, H., Hoover, W.: {\it Non equilibrium molecular
dynamics of a classical fluid}, in "Molecular Liquids: new perspectives
in Physics and chemistry", ed. J. Teixeira-Dias, Kluwier Academic
Publishers, p. 527--547, 1992.

\item{[R1]} Ruelle, D.: {\it Chaotic motions and strange
attractors}, Lezioni Lincee, notes by S. Isola, Accademia Nazionale dei
Lincei, Cambridge University Press, 1989; see also: Ruelle, D.: {\it
Measures describing a turbulent flow}, Annals of the New York Academy of
Sciences, {\bf 357}, 1--9, 1980.  For more technical expositions see
Ruelle, D.: {\it Ergodic theory of differentiable dynamical systems},
Publications Math\'emathiques de l' IHES, {\bf 50}, 275--306, 1980.

\item{[R2]} Ruelle, D.: {\it A measure associated with axiom A
attractors}, American Journal of Mathematics, {\bf98}, 619--654, 1976.

\item{[R3]} Ruelle, D.: {\it Thermodynamic formalism}, Encyclopedia of
Mathematics, vol. 5, Addison-Wesley, 1978.

\item{[R4]} Ruelle, D. {\it Ergodic theory of differentiable dynamical
systems},
Publications Mathematiques de l' IHES, {\bf 50}, 27--58, 1979.

\item{[R5]} Ruelle. D.: {\it Statistical mechanics of a one dimensional
lattice gas}, Communications in Mathematical Physics, {\bf 9}, 267--278,
1968.

\item{[S1]} Sinai, Y.: {\it Dynamical systems with elastic reflections.
Ergodic properties of dispersing billards},
Russian Mathematical Surveys, {\bf 25}, 137--189, 1970.

\item{[S2]} Sinai, Y.: {\it Topics in ergodic theory}, Princeton
University Press, 1994. Sinai, Y.: {\it Markov partitions and
$C$-diffeomorphisms}, Functional Analysis and Applications, {\bf2},
64--89, 1968, n.1 (engl. p. 61); and Sinai, Y.: {\it Construction of Markov
partitions}, Functional analysis and Applications, {\bf2}, 70--80, 1968,
n.2 (engl. p. 245).

\item{[Sm]} Smale, D.: {\it Differentiable dynamical systems}, Bulletin of
the american mathematical society, {\bf 73}, 747--817, 1965.

\item{[SEM]} Sarman, S., Evans, D., Morriss, G.: {\it Conjugate pairing
rule and thermal transport coefficients}, Physical Review, {\bf 45A},
2233--2242, 1992.

\item{[T]} Thomson, W.: {\it The kinetic theory of the dissipation of
energy}, in the collected "Mathematics and Physics papers", vol. V, p.
11-20, 1911, Cambridge University Press, from the 1874 original.

\ciao
\bye